\documentclass[a4paper,aps,prb,floats,twocolumn]{revtex4}
\usepackage{overpic}
\usepackage{graphicx, latexsym, verbatim}
\usepackage{graphics}
\usepackage{amssymb, amsmath}
\usepackage[ansinew]{inputenc}
\usepackage{color}
\usepackage{multirow}
\usepackage{epstopdf}
\usepackage{tikz}
\pgfrealjobname{modeVolumeControversy}

\usepackage{pgfplots}
\usepgfplotslibrary{groupplots}
\usepgfplotslibrary{colormaps}
\pgfplotsset{compat=newest}

\pgfplotsset{plot coordinates/math parser=false}
\newlength\figureheight
\newlength\figurewidth
\pgfplotsset{/pgfplots/colormap/.code 2 args={
	\pgfplotscreatecolormap{#1}{#2}
	\pgfkeysalso{/pgfplots/colormap name={#1}}%
}}

\pgfplotsset{
tick label style={font=\footnotesize},
label style={font=\footnotesize},
legend style={font=\footnotesize}
}

\definecolor{darkblue}{rgb}{0,0,0.55}
\definecolor{comfrey}{rgb}{0.85,0.85,0.85}
\usepackage{rotating}
\newcommand{\ud}{\mathrm{d}}
\newcommand{\mr}{\mathbf{r}}

\newcommand{\mft}{\tilde{\mathbf{f}}}
\newcommand{\tlo}{\tilde{\omega}}

\definecolor{myBlue}{rgb}{0.0430,0.5156,0.7773}

\newcommand{\rev}[1]{{\color{black}#1}}
\newcommand{\revTwo}[1]{{\color{black}#1}}
\newcommand{\revrev}[1]{{\color{black}#1}}
\newcommand{\revrevrev}[1]{{\color{black}#1}}
\newcommand{\comm}[1]{{}}
\newcommand{\commMul}[1]{{}}

\newcommand{\takeout}[1]{{}}

\begin{document}

% reduce the space around align equations
\abovedisplayskip=7pt
\abovedisplayshortskip=0pt
\belowdisplayskip=7pt
\belowdisplayshortskip=7pt

\bibliographystyle{thesis_bibliographystyle}
% Bibliography styles that can be used instead of prsty are abbrv, alpha, plain and unsrt

%Title of paper
\title{Normalization of quasinormal modes in leaky optical cavities and plasmonic resonators }% of spherically symmetric systems}
%\title{On the effective volume of quasinormal modes \comm{in optical cavities and plasmonic resonators} }% of spherically symmetric systems}
\author{Philip Tr\o st Kristensen}
\affiliation{Institut f\"ur Physik, Humboldt Universit\"at zu Berlin, 12489 Berlin, Germany}
\author{Rong-Chun Ge}
\affiliation{Department of Physics, Queen's University, Kingston, Ontario K7L 3N6, Canada}
\author{Stephen Hughes}
\affiliation{Department of Physics, Queen's University, Kingston, Ontario K7L 3N6, Canada}

\date{\today}

%\maketitle must follow title, authors, abstract, \pacs, and \keywords

\begin{abstract}
%\comm{needs some work:}
We discuss three formally different formulas for %elaborate on the various
normalization of %formulas for
quasinormal modes currently in %current
use for modeling optical cavities and plasmonic resonators and show that they are complementary and provide the same result.
% they are ultimately
%discuss how they derive from independent descriptions of the local optical field in terms of quasinormal modes.  For a number of example problems, we show explicitly that the three formulations are ultimately equal and provide the same results.
Regardless of the formula used for normalization, one can use the norm to define an effective mode volume for use in Purcell factor calculations.\\ \\  \end{abstract}

\maketitle

%This note addresses a recent concern about the correctness as well as the %practical evaluation of normalization integrals for quasinormal modes (QNMs) %of optical resonators. The issue was noted by Christophe Sauvan (in his %talk at SPIE Optics + Photonics 2014), who pointed out that the integrand %in the normalization integral of Lai~\emph{et al.}~\cite{Lai_PRA_41_5187_1990} %grows without bounds for sufficiently large distances from the cavity. In %a recent preprint, the issue is emphasized by Muljarov \emph{et al.}~\cite{Muljarov_arXiv_1409_6877_2014}, %who also point out that an alternative formulation for the normalization %integrand has been derived earlier~\cite{Muljarov_EPL_92_50010_2010}.

\section{Introduction}
%\rev{Something about the general use of QNMs for open resonators and a summary of the recent works, including Lalanne's, Leung's, and a timeline}
Optical cavities and plasmonic resonators enable tight localization of electromagnetic fields to length scales on the order of the wavelength or shorter. In addition, the resonant nature of these material systems suggest a discrete nature of the %that the
possible electromagnetic waves, %should be discrete,
similar to the bound states of the hydrogen atom. The analogy is somewhat deceptive because of the inherently leaky nature of electromagnetic  resonators, which leads to % are inherently leaky, resulting in
a continuous dissipation of energy via radiation or absorption in the material. Nevertheless, distinct resonances %the spectral locations of the resonances
show up as peaks in spectra from scattering calculations or measurements, for example, in which the width of the peak is related to a finite lifetime due to the dissipation of energy; % via the imaginary part of the frequency;
the associated scattered fields often show signatures of an underlying modal structure. Throughout the literature, these resonances are frequently referred to as ``modes'', but an explicit mathematical definition of the modes is rarely given, and this may cause both conceptual and practical difficulties when using the modes in calculations. %sometimes missing. %the underlying modal structure due to the QNMs.
\begin{figure}[tb]
\flushright
\includegraphics{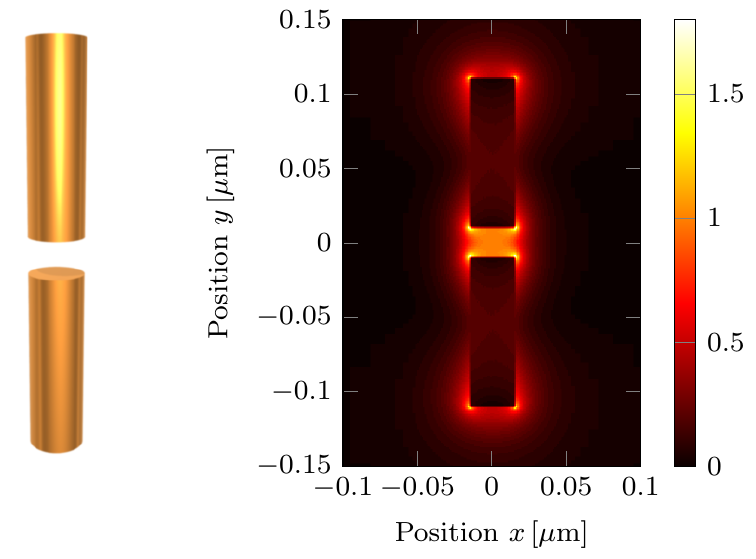}
%\beginpgfgraphicnamed{fig1}
%\raisebox{.95cm}{\begin{overpic}[height=0.525\columnwidth]{nanoRodGoldDimer.png}
%%%\put(31,-5){Position $x/a$}
%%%\put(13,26.5){(a)}
%%%\put(-12,0){\begin{sideways}{Numerical Integral}\end{sideways}}
%%\put(-7,14){\begin{sideways}{Position $y/a$}\end{sideways}}
%%\put(33,-6){Position $x/a$}
%\end{overpic}}\qquad\quad
%%\includegraphics[height=0.525\columnwidth]{nanoRodGoldDimer.png}
%\begin{tikzpicture}
%\begin{axis}[
%	width=0.35\columnwidth,
%	height=0.525\columnwidth,
%	enlargelimits=false,
%	axis on top,
%	scale only axis,
%	colormap name=hotAsHell,
%	point meta min=0,
%	point meta max=1.8,
%	colorbar,
%	colorbar/width=0.025\columnwidth,
%%	colorbar/height=0.6\columnwidth,
%%	colorbar style={
%%		height=0.6\columnwidth,
%%		point meta min=-1,
%%		point meta max=1,
%%		ytick={-1,-0.5,...,1}},
%	xlabel={Position $x\,[\mu$m]},
%	ylabel={Position $y\,[\mu$m]},
%	ytick={-0.15,-0.1, ..., 0.15},
%	ticklabel style={
%		/pgf/number format/precision=2,
%		/pgf/number format/fixed,
%		},
%	xtick={-0.1,-0.05, ..., 0.1},
%]
%\addplot graphics
%[xmin=-0.1,xmax=0.1,ymin=-0.15,ymax=0.15]
%{plasmonDimer_absField_v3};
%\end{axis}
%\end{tikzpicture}%
%\endpgfgraphicnamed
%\\[-2.5mm]
\caption{\label{Fig:plasmonDimer_absField}(Color
  online) Left: Illustration of a \rev{gold} nanorod dimer. Right: Absolute value of the fundamental QNM $\mft_\text{c}(\mr)$ in the plane $z=0$. The QNM is scaled to unity at the position $\mr_\text{c}=(0,0,0)$ in the center between the two rods. % atcavity center $\mr_\text{c}=(0,0)$.
%\comm{[Maybe include here a spectrum and refer to the figure in the first paragraph.]}
}
\end{figure}

%Mathematically,
In computational nanophotonics, the optical or plasmonic response of resonant systems are often modeled by use of time-domain calculations with Perfectly Matched Layers (PMLs) and a run-time Fourier transform to extract the modal field distribution. This technique has been very successful for modeling high quality optical cavities for which the spectral resonances are well separated. In particular, the modes so calculated have been used for estimates of the spontaneous emission rate enhancement via the Purcell formula\cite{Purcell_PR_69_681_1946}, from which the spontaneous emission rate is expected to scale with the ratio of the cavity quality factor $Q$ to the effective mode volume $V_\text{eff}$. The effective mode volume describes the degree of light localization in the cavity and has a clear definition for closed resonators in terms of the integrated energy density of light in the mode. The definition, however, cannot be directly extended to dissipative systems for which the mode extends over all space, and this has conceptual consequences for application of the Purcell formula which manifestly pertains to dissipative systems with a finite $Q$ value. % %\cite{Kristensen_OL_37_1649_2012}.
%in fact so-called quasinormal modes (QNMs), which appear as the discrete solutions to the wave equation that satisfy a radiation condition compatible with the propagation of light away from the resonator.
%
Theoretically, the response of dissipative resonant structures can be %very conveniently modeled and
understood in terms of so-called quasinormal modes~\cite{Ching_1996, Ching_RevModPhys_70_1545_1998, Kristensen_ACSphot_1_2_2014} (QNMs), which appear as discrete solutions to the wave equation and satisfy a radiation condition compatible with the propagation of light away from the resonator. Figure~\ref{Fig:plasmonDimer_absField} shows an example of a plasmonic resonator, in the form of a nanorod dimer, and the field profile of the fundamental QNM in the dimer. The use of PMLs %Perfectly Matched Layers (PMLs)
is one possible realization of a radiation condition for resonators in homogeneous media, wherefore it was pointed out in Ref.~\onlinecite{Kristensen_OL_37_1649_2012} %remarked
that the optical cavity modes commonly calculated by time-domain techniques are exactly the QNMs, and that the Purcell formula can be derived within a QNM framework such as developed in Refs.~\onlinecite{Lai_PRA_41_5187_1990, Leung_PRA_49_3057_1994, Leung_JOSAB_13_805_1996, Lee_JOSAB_16_1409_1999, Lee_JOSAB_16_1418_1999}, for example; this leads naturally to a definition of the effective mode volume which is compatible with the leaky nature of the modes. The effective mode volume is intimately connected with the normalization of the QNMs, and two alternative formulations of the effective mode volume %resulted in different formulations of the effective mode volume
were derived in Refs.~\onlinecite{Sauvan_PRL_110_237401_2013} and~\onlinecite{Muljarov_arXiv_1409_6877_2014}\revrev{, respectively,} based on different formulations of the norm. %Regardless of the formula used for normalization, however, one can use the norm to define an effective mode volume for use in Purcell factor calculations.
In this article we show how the three different normalization integrals are closely related %and complementary
and provide the same result. %Below we discuss the three different normalization formulas in detail and show how they are related.
In essence, we show that the differences in the formulas can be understood as alternative regularization procedures for an inherently ill-behaved %, yet formally convergent,
integral. %We elaborate on the formal need for a regularization of the normalization integral
%The divergent behavior of the QNMs lead to the formal need for a regularization of the integrals, and we show that
Although we illustrate the formal connection between the three formulations, it is worth emphasizing that they were derived by use of %independent methods.
%he three formulations of the norm were derived using
very different approaches in Refs.~\onlinecite{Lai_PRA_41_5187_1990}, \onlinecite{Sauvan_PRL_110_237401_2013} and \onlinecite{Muljarov_EPL_92_50010_2010}. The equivalence of the formulas, therefore, should be regarded as a profound strength of the entire modeling framework for localized electromagnetic resonators based on QNMs.

%T
The article is organized as follows: In section~\ref{Sec:normIntegrals} we introduce the three formulations of the QNM norm and discuss how they relate to each other. Section~\ref{Sec:Examples} provides explicit example calculations to illustrate the practical evaluation as well as the equivalence of the norms. Section \ref{Sec:Conclusions} summarizes and concludes the article.

\section{Normalization integrals}
\label{Sec:normIntegrals}
We consider electric field QNMs $\mft_\mu(\mr)$ of resonators embedded in a homogeneous background defined as the eigenmodes of Maxwell's wave equation
\begin{align}
\nabla\times\nabla\times\mft_\mu(\mr)-\tilde{k}_\mu^2 %\left(\frac{\tlo_\mu}{\text{c}}\right)^2
\epsilon_\text{r}(\mr,\tlo_\mu)\mft_\mu(\mr)=0,
\end{align}
in which $\tilde{k}_\mu=\tlo_\mu/\text{c}$ is the ratio of the angular resonance frequency to the speed of light and $\epsilon_\text{r}(\mr,\tlo_\mu)$ is the relative permittivity.
\rev{Because of the leaky nature of the resonators, the \revTwo{differential equation should be augmented by a radiation condition to only allow outgoing wave solutions at large distances.} %QNMs should obey a radiation condition.
For the case of a homogeneous environment, we argue that this radiation condition is formally the}
%a radiation condition}
 %$\tilde{k}_\mu=n_\text{B}\omega_\mu/\text{c}$ is the magnitude of the wave vector in the homogeneous background material with refractive index $n_\text{B}$.
%subject to the
Silver-M\"uller radiation condition~\cite{Martin_MultipleScattering} \revTwo{in the form %which may be written as
\begin{align}
\frac{\mathbf{r}}{r}
%\mathbf{r}
\times\nabla\times\mft_\mu(\mr) \rightarrow -\text{i}n_\text{B}\tilde{k}_\mu\mft_\mu(\mr) \quad\text{as}\;r\rightarrow\infty,
\label{Eq:SilverMullerCond}
\end{align}
where} $n_\text{B}$ is the refractive index of the homogeneous background material. % as $r=|\mr|\rightarrow\infty$.
\revTwo{The Silver-M\"uller radiation condition is commonly imposed to correctly model scattered fields. Therefore, we argue that this is also the radiation condition for the QNMs, which can be considered solutions to the scattering problem with no source.} %
\rev{Because it is defined only in the limit of infinite distances, Eq.~(\ref{Eq:SilverMullerCond}) is rarely useful in} \revTwo{numerical} \rev{calculations of QNMs. Instead, one can use PMLs to mimic the radiation condition by removing reflections from the calculation domain boundary, %\cite{
or one can use calculation methods based on the Green tensor which manifestly respects the radiation condition.} % The QNMs can be regarded as the solutions to a scattering problem with no incident field.} %\rev{Although the Silver-M\"uller radiation condition appears formally to be the correct choice of radiation condition, }
As a result of the radiation condition, the differential equation problem becomes non-Hermitian, wherefore a number of well-known text book results for Hermitian eigenvalue problems do not apply. In particular, the discrete resonance frequencies $\tlo_\mu=\omega_\mu-\text{i}\gamma_\mu$ are complex with a negative imaginary part, from which the $Q$ value can be calculated as $Q_\mu=\omega_\mu/2\gamma_\mu$. %\revrev{Although we shall work primarily with electric field QNMs as defined above, it is immediately clear that for each electric field QNM there is an associated magnetic field QNM, %satisfying 
%\begin{align}
%\tilde{\mathbf{g}}_\mu(\mr) =-\frac{\text{i}}{\tlo_\mu}\nabla\times\mft_\mu(\mr)
%\end{align}
%$\tilde{\mathbf{g}}_\mu(\mr) =\nabla\times\mft_\mu(\mr)/\text{i}\tlo_\mu\mu_0\mu_\text{r}(\mr,\tlo_\mu)$, where $\mu_0$ and $\mu_\text{r}(\mr,\omega)$ denote the permittivity of free space and the (possibly dispersive) relative permeability, respectively.}
The complex resonance frequencies in combination with the radiation condition leads to a divergent behavior of the QNMs as a function of distance from the resonator and makes the %causes the normalization to become %wherefore the normalization becomes
normalization non-trivial. As discussed in the introduction, the problem of normalization for QNMs has been addressed by at least three different methods leading to three different formulas that we discuss in detail below. Before moving on, however, it is worth elaborating a bit on the connection between the mathematical concept of the norm, and the more physical concept of mode volume.

Regardless of the formula used for normalization, one can use the norm to define a generalized effective mode volume for the QNM of interest (denoted by $\mu=\text{c}$) as\cite{Kristensen_OL_37_1649_2012} %$\mft_\text{c}(\mr)$ as
\begin{align}
v_\text{c} = \frac{\langle\langle\mft_\text{c}|\mft_\text{c}\rangle\rangle}{\epsilon_\text{r}(\mr_\text{c})\mft^2_\text{c}(\mr_\text{c})},
\label{Eq:v_Q_def}
\end{align}
where $\langle\langle\mft_\text{c}|\mft_\text{c}\rangle\rangle$ denotes the QNM norm (in any formulation compatible with the \revrev{divergent behavior %properties 
of the QNMs at large distances}) %, $\mr_\text{c}$ denotes the center of the cavity, 
and $\mft^2_\text{c}(\mr_\text{c})=\mft_\text{c}(\mr_\text{c})\cdot\mft_\text{c}(\mr_\text{c})$. The position $\mr_\text{c}$ denotes a point of special interest to the given resonator. % and for which the ...
For optical cavities, this will often be the center of the cavity in which the field is largest, and the \revTwo{generalized effective mode volume then} gives a measure for the maximum \revTwo{light-matter interaction} %spontaneous emission enhancement
that one can obtain \revTwo{in} %for an emitter in
the cavity. For general resonators %In general cavities
there may be no unique way of defining the center, or it may be that the mode vanishes at this point. % Typically, however, there will be a  $\mr_\text{c}$
For plasmonic resonators, in particular, the field may be several orders of magnitude larger at the edges than within the material. Despite the large field values, however, the edges are often uninteresting for \revTwo{light-matter interaction} %spontaneous emission enhancement
because of enormous non-radiative decay and quenching. In general, therefore, we specify $\mr_\text{c}$ explicitly for each resonator.
%; one can then either define a position-dependent mode volume
%may be advantageous to define $\mr_\text{c}$
%Therefore, the choice of $\mr_\text{c}$ is somewhat arbitrary in the general case, but often there will be a position of special interest for which one can define the effective mode volume.
% \revTwo{plus add discussion about approximation, and for which the $Q$ value is sufficiently high, so that dacay rate is much smaller than the angular frequency, so that $\tlo_\text{c}/\omega_\text{c}\approx1$}.
The generalized effective mode volume is complex in general, and is related to the effective mode volume $V_\text{eff}$ of the Purcell formula %\revTwo{Purcell formula
%\begin{align}
%F_\text{P} = \frac{3}{4\pi^2}\left(\frac{\lambda_\text{c}}{n_\text{c}}\right)^3\left(\frac{Q}{V_\text{eff}}\right),
%\label{Eq:PurcellFactor}
%\end{align}
as
\begin{align}
\frac{1}{V_\text{eff}} = \text{Re}\left\{\frac{1}{v_\text{c}}\right\}.
\label{Eq_V_eff_def}
\end{align}
This definition follows directly and quite naturally from a derivation of the original Purcell formula in the framework of QNMs, either using a Green tensor approach\cite{Kristensen_ACSphot_1_2_2014, Kristensen_OL_37_1649_2012, Muljarov_arXiv_1409_6877_2014} or a %more traditional
mode expansion formulation\cite{Sauvan_PRL_110_237401_2013}. It follows from these derivations, that %Importantly,
the usefulness of the effective mode volume %\rev{in Eq.~(\ref{Eq_V_eff_def})}
for Purcell factor calculations is restricted to material systems or frequency ranges for which a single mode dominates the response. \revTwo{Moreover, the use of Eq.~(\ref{Eq_V_eff_def}) in Purcell factor calculations is valid in the limit $\tlo_\text{c}/\omega_\text{c}\approx1$ which may not hold for very low-$Q$ optical cavities or plasmonic resonators. In cases where it does not hold, one can still recover the Purcell formula by redefining either Eq.~(\ref{Eq:v_Q_def}) or (\ref{Eq_V_eff_def}), but \revrev{in practice it is often easier and more convenient} %it may be easier 
to calculate the spontaneous emission enhancement directly from a QNM expansion of the Green tensor. %\revrev{enhanced spontaneous emission rate $\Gamma$ relative to the corresponding rate in a homogeneous medium, $\Gamma_\text{B}$, directly from a QNM expansion of the Green tensor $\mG(\mr,\mr',\omega)$ as
%\begin{align}
%\frac{\Gamma(\mr_\text{c},\omega)}{\Gamma_\text{B}(\omega)} = \frac{\text{Im}\left\{\mG(\mr_\text{c},\mr_\text{c},\omega)\right\}}{\text{Im}\left\{\mG_\text{B}(\mr_\text{c},\mr_\text{c},\omega)\right\}},
%\label{Eq:F_P_from_G}
%\end{align}
%where $\mG_\text{B}(\mr,\mr',\omega)$ is the Green tensor for the homogeneous medium, and  $\text{Im}\left\{\mG_\text{B}(\mr_\text{c},\mr_\text{c},\omega)\right\}=n_\text{B}\omega/6\pi\text{c}$.} 
For plasmonic structures, the %Moreover, the
usefulness of the original Purcell formula %for plasmonic structures
is further limited by non-radiative decay channels at positions very close to the material~\cite{Koenderink_OL_35_4208_2010,Ge_arXiv_1312.2939_2013}.} %Finally, it is worth pointing out that in}

\subsection{Normalization by Lai \emph{et al.}}
\label{Sec:LaiNorm}
\revrevrev{Lai \emph{et al.} introduced a normalization for QNMs, which was applied to the study of spherically symmetric and non-dispersive material systems in Ref.~\onlinecite{Lai_PRA_41_5187_1990}. %For spherically symmetric \revrevrev{material systems} %resonators  and non-dispersive materials, Lai \emph{et al.} introduced % the normalization~\cite{Lai_PRA_41_5187_1990}
%\revrevrev{a normalization for QNMs in Ref.~\onlinecite{Lai_PRA_41_5187_1990}. 
Rewriting slightly, we express this norm as} %in a form which is independent of coordinate system as}
\begin{align}
\langle\langle\mft_\mu|\mft_\mu\rangle\rangle_\text{Lai} = &\lim_{V\rightarrow\infty}\Big\{\int_V \epsilon_\text{r}(\mr)\mft_\mu(\mr)\cdot\mft_\mu(\mr)\ud V \nonumber \\
&+\text{i}\frac{n_\text{B}}{2\tilde{k}_\mu} \int_{\partial V} \mft_\mu(\mr)\cdot\mft_\mu(\mr)\ud A\Big\},
\label{Eq:LaiNorm}
\end{align}
where $V$ is a spherical volume with boundary $\partial V$ and $\ud V$ and $\ud A$ are differential volume and area elements, respectively. %Historically,
The normalization in Ref.~\onlinecite{Lai_PRA_41_5187_1990} %Eq.~(\ref{Eq:LaiNorm}) 
appears to date back to early work by Zel'dovich on the theory of unstable states\cite{Zeldovich_SPJ_12_542_1961}. It %This normalization
was later adopted\revrevrev{, in the form of Eq.~(\ref{Eq:LaiNorm}), to introduce a generalized effective mode volume} %by Kristensen, Van Vlack and Hughes,
for %general, 
leaky optical cavities~\cite{Kristensen_OL_37_1649_2012}. The formula can be extended to dispersive resonators in homogeneous and non-dispersive surroundings by the substitution\cite{Leung_PRA_49_3057_1994} $\epsilon_\text{r}(\mr)\rightarrow\sigma(\mr,\tlo_\mu)$, where
\begin{align}
%\sigma(\mr,\omega) = \frac{1}{2\omega}\frac{\partial}{\partial\omega}\big(\omega^2\epsilon_\text{r}(\mr,\omega)\big).
\sigma(\mr,\omega) = \frac{1}{2\omega}\partial_\omega\big(\omega^2\epsilon_\text{r}(\mr,\omega)\big).
\label{Eq:sigmaDef}
\end{align}
% and related to the effective mode volume of the Purcell formula. %actor.
%, where we introduced an approach for computing the exact generalized (complex) effective mode volume and Purcell factor for an open cavity system.}
%\commMul{Muljarov \emph{et al.}~\cite{Muljarov_arXiv_1409_6877_2014} refers to Eq.~(\ref{Eq:LaiNorm}) as the ``Normalization by Kristensen \emph{et al.}''. It is worth pointing out, however, that Eq.~(\ref{Eq:LaiNorm}) was not derived in Ref.~\onlinecite{Kristensen_OL_37_1649_2012}, %[OL, 37 1649 (2013)], although we do accept responsibility for applying it to non-spherically symmetric resonators. %, and for using it to compute the effective mode volume and the Purcell factor. Therefore, although it is very flattering to be credited with this nice normalization, we find it to be rather inappropriate to  refer to Eq.~(\ref{Eq:LaiNorm}) as the ``Normalization by Kristensen \emph{et al.}'', especially when applied to a spherical material system as in Ref.~\onlinecite{Muljarov_arXiv_1409_6877_2014}.}

\revrevrev{The norm in %In rewriting the norm of Lai \emph{et al.}\cite{Lai_PRA_41_5187_1990} in the form of 
Eq.~(\ref{Eq:LaiNorm}) suffers from a pathological deficiency in the case of degenerate modes of high-symmetry material systems. For spherical resonators, for example, one can choose the azimuthal dependence of the QNMs to be of the form $\exp\{\pm\text{i}m\varphi\}$, for integer values of $m$. This choice will lead to a vanishing norm. %, so that Eq.~(\ref{Eq:LaiNorm}) acts as a semi-norm in these cases. 
As discussed in Ref.~\onlinecite{Lai_PRA_41_5187_1990}, it may therefore be more appropriate to formulate the norm in terms of the adjoint fields which, for spherical systems, is obtained by complex conjugation of the angular dependence only\cite{Lai_PRA_41_5187_1990}. It is clear, however, that any perturbative breaking of the symmetry will lift the degeneracy and lead to well-defined QNMs for which the norm applies immediately. Therefore, in such cases where the norm vanishes, one can always choose linear combinations of the degenerate QNMs for which the norm applies as written in Eq.~(\ref{Eq:LaiNorm}). In the case of the spherical resonator, for example, one can choose the azimuthal dependencies of the QNMs to be of the form $\exp\{\text{i}m\varphi\}\pm\exp\{-\text{i}m\varphi\}$; in this case Eq.~(\ref{Eq:LaiNorm}) applies directly.

%As a physical motivated example of such a linear combination, we consider the case of the spherical resonator with a perturbative change in the refractive index of the form $\Delta\epsilon_\text{r}(\varphi)$. In the limit of vanishing perturbation, this approach fixes the azimuthal dependencies of the QNMs to be of the form $\exp\{\text{i}m\varphi\}\pm\exp\{-\text{i}m\varphi\}$ for which Eq.~(\ref{Eq:LaiNorm}) applies directly.
}

For any resonator, \revrev{we} can evaluate the norm in Eq.~(\ref{Eq:LaiNorm}) by first integrating over a spherical domain $V_0$ of radius $R_0$ completely enclosing the resonator; this integral is well behaved and %clearly, this integral 
can be easily evaluated. \revrev{Next, for} %For 
the region $r>R_0$, we can write the QNMs in terms of spherical Hankel functions $h_l(n_\text{B}\tilde{k}_\mu r)$ and vector spherical harmonics\cite{Barrera_EJP_6_287_1985}. % (see Appendix~\ref{App:VSH}) as
%\begin{align}
%\mft(\mr) = \sum_{lm}h_l(kr)\mathbf{q}_{lm}\cdot\mathbf{Q}_{lm}(\theta,\varphi)
%\end{align}
%where $\mathbf{q}_{lm} = [V_{lm}^r,E_{lm}^{(1)},E_{lm}^{(2)}]^\text{T}$, and
%\begin{align}
%\mathbf{Q}_{lm}(\theta,\varphi) = \left[ \begin{array}{c}
%\mathbf{Y}_{lm}(\theta,\varphi)  \\
%\mathbf{\Psi}_{lm}(\theta,\varphi)\\
%\mathbf{\Phi}_{lm}(\theta,\varphi)\\
%\end{array} \right].
%\end{align}
%\cite{Barrera_EJP_6_287_1985}, %
%\begin{align}
%\mathbf{Q}_{lm}(\theta,\varphi) = \left[\mathbf{Y}_{lm}(\theta,\varphi),\mathbf{\Psi}_{lm}(\theta,\varphi),\mathbf{\Phi}_{lm}(\theta,\varphi)\right],
%\end{align}
\revrev{Therefore,} %wherefore 
the norm may be rewritten in the form %integral reduces to a sum of terms of the form
%\begin{align}
%I_\text{L}^{r>R_0} = \int_\Omega \sum_{lmn} I_l\left({q}^n_{lm}{Q}^n_{lm}(\theta,\varphi)\right)^2  \ud\Omega
%\end{align}
%where $\Omega$ denotes the sphere and  $\ud\Omega=\sin\theta\ud\theta\ud\varphi $.
%
\begin{align}
\langle\langle\mft_\mu|\mft_\mu\rangle\rangle_\text{Lai} = \int_{V_0} \epsilon_\text{r}(\mr)\mft_\mu(\mr)\cdot\mft_\mu(\mr) \ud V + \sum_{lm} I^r_l I^\Omega_{lm},
\label{Eq:LaiNorm_sphericalForm}
\end{align}
in which $I^r_l = \lim_{R\rightarrow\infty}\{\hat{I}^r_l(R)\}$ with
\begin{align}
\hat{I}^r_l(R) =\!\int_{R_0}^R\! n_\text{B}^2h_l^2(n_\text{B}\tilde{k}_\mu r)r^2\ud r +\text{i}\frac{n_\text{B}}{2\tilde{k}_\mu}h_l^2(n_\text{B}\tilde{k}_\mu R)R^2,
\label{Eq:Ioutside}
\end{align}
and $I^\Omega_{lm}$ includes %the expansion coefficients and
the angular integration of the vector spherical harmonics, which is independent of $R$ (and finite). 
\revrev{Here, and in the following, we use linear combinations of vector spherical harmonics of the form $\tilde{\mathbf{P}}_{lm}^\pm = \mathbf{P}_{lm}\pm\mathbf{P}^*_{lm}$, where $\mathbf{P}_{lm}$ is any vector spherical harmonic of order $(l,m)$ in the formulation of Ref.~\onlinecite{Barrera_EJP_6_287_1985}.} % This ensures that the vector spherical harmonics are orthogonal in the sense of Eq.~(\ref{})  \cite{Barrera_EJP_6_287_1985}, %the sphere}
%\begin{align}
%I^\Omega_{lm} = \int_\Omega\left(\right) \ud\Omega
%\end{align}
%\begin{align}
%I_\Omega = \int_\Omega\mathbf{X}_{lm}(\theta,\varphi)\ud\Omega
%\end{align}
%is independent of $r$ (and finite).
The argument, originally by Lai \emph{et al.}~\cite{Lai_PRA_41_5187_1990}, is now, that because of the outgoing wave nature of the Hankel functions at large distances, the increase in the volume integral is exactly balanced by the additional surface integral. For large arguments, the spherical Hankel functions tend to the limiting form% \comm{[Ref]}
\begin{align}
h_l(z)\rightarrow\text{e}^{-\text{i}\pi(l+1)/2}\frac{1}{z}\text{e}^{\text{i}z},\quad z\rightarrow\infty,
\label{Eq:HankelLimForm}
\end{align}
so that inserting in Eq.~(\ref{Eq:Ioutside}) and differentiating with respect to $R$ we find
\begin{align}
%\frac{\partial}{\partial R}
\partial_R\hat{I}^r_l(R) &= (-1)^{(l+1)}\frac{1}{\tilde{k}_\mu^2}\Big\{ \text{e}^{2\text{i}n_\text{B}\tilde{k}_\mu R}+\frac{\text{i}}{2n_\text{B}\tilde{k}_\mu}\partial_R%\frac{\partial}{\partial R}
\Big(\text{e}^{2\text{i}n_\text{B}\tilde{k}_\mu R}\Big) \Big\}\nonumber\\
&= 0,
\label{Eq:varIoutside}
\end{align}
suggesting that %\rev{Eq.~(\ref{Eq:LaiNorm}) ``has an unambiguous meaning independnt of $R$''\cite{Lai_PRA_41_5187_1990}.}
one can assign a well-defined value to the norm as $R\rightarrow\infty$.

In practice, direct application of Eq.~(\ref{Eq:LaiNorm}) leads to an integral that seems to quickly converge towards a finite value, but in fact oscillates about this value with an amplitude that eventually starts to grow (exponentially) with the distance, albeit slowly compared to the %typical
length scales in typical %most
calculations. This was noted %by two of us
in Ref.~\onlinecite{Kristensen_OL_37_1649_2012}, where the oscillations were observed only for the cavity with the lowest quality factor ($Q\approx16$). The source of the oscillating integral can be traced back to the complex resonance frequency, which means that the wavenumber $\tilde{k}_\mu$ %$k=k_\text{R}-\text{i}\gamma_k$
is complex as well. Evaluation of Eq.~(\ref{Eq:Ioutside}) then % the integral
leads to residual radius dependent terms of the form
\begin{align}
f_\text{Lai}^\text{res}(R) = \sum_n\frac{\text{1}}{P_n(R)}\text{e}^{2\text{i}n_\text{B}\tilde{k}_\mu R},
\label{Eq:f_res_Lai_general}
\end{align}
where $P_n(R)$ denotes general polynomials in $R$ of order $n>1$; the surface term in Eq.~(\ref{Eq:Ioutside}) cancels the first order term. %For complex wave vectors $k$, it is clear that the terms in Eq.~(\ref{Eq:f_res_Lai_general}) must grow beyond bounds
Thus, while Eqs.~(\ref{Eq:HankelLimForm}) and (\ref{Eq:varIoutside}) appear to be formally correct also for complex arguments, the limit $R\rightarrow\infty$ in practice leads to a position dependent phase difference between the Hankel function and its limiting form, which makes the limit non-trivial to perform along the real axis. In principle, however, \revrev{it is possible to} %one can 
regularize the integral using coordinate transforms; \rev{a technique that has \revrev{also} been also for normalization of leaky modes in waveguides\cite{SnyderLoveBook}. Formally, we can rewrite the} integral as
\begin{align}
 I_{l}^r  &= \lim_{R\rightarrow\infty}\Big\{\int_{R_0}^Rn_\text{B}^2 h_l^2(n_\text{B}\tilde{k}_\mu r)r^2\ud r + \text{i}\frac{n_\text{B}}{2\tilde{k}_\mu}h_l^2(n_\text{B}\tilde{k}_\mu R)R^2 \nonumber\\
&\quad\qquad - \int_{R_0}^R \partial_r f_\text{Lai}^\text{res}(r)\ud r + \int_{R_0}^R \partial_r f_\text{Lai}^\text{res}(r)\ud r \Big\} \nonumber \\
&= F_l(R_0) + f_\text{Lai}^\text{res}(R_0)+ \lim_{R\rightarrow\infty}\int_{R_0}^R \partial_r f_\text{Lai}^\text{res}(r)\ud r,
\label{Eq_I_r_l}
\end{align}
where
\begin{align}
F_l(r) %&= \int h_l^2(kr)r^2\ud r \\
&= -\frac{r^3n_\text{B}^2}{2}\left( h_l^2(n_\text{B}\tilde{k}_\mu r) - h_{l-1}(n_\text{B}\tilde{k}_\mu r)h_{l+1}(n_\text{B}\tilde{k}_\mu r) \right)
\end{align}
%$F_l(R_0)$
is minus the antiderivative of $f_l(r) = n_\text{B}^2h_l^2(n_\text{B}\tilde{k}_\mu r)r^2$. In the last integral, we can now change the integration contour by writing $r=R_0+iz$ to find  $I_l^r= F_l(R_0)$. For the case of a spherical resonator, we provide an example of this approach in section \ref{Sec:Spheres}. Obviously, this regularization approach is rather cumbersome unless the expansion in spherical wave functions is explicitly known; also, it is rarely necessary in practice. Although the integral oscillates as a function of calculation domain size, the associated uncertainty in the calculated norm is much lower than the general numerical uncertainty for most QNM calculations. We elaborate on this issue in sections \ref{Sec:pcCavity} and \ref{Sec:plasmonicRod}.

%%\subsection*{Discussion}
%Comparing Eqs.~(\ref{Eq:LaiNorm}), (\ref{Eq:SauvanNorm}) and (\ref{Eq:MuljarovNorm}), we can understand the three formulations of the norm as arising from three different approaches to the regularization of the integral
%
%It is worth noting, that from the above discussion it is clear that the integral
%\begin{align}
%\int\epsilon_\text{r}(\mr)\mft_\mu(\mr)\mft_\mu(\mr)\ud V
%\end{align}
%can be regularized immediately by a complex coordinate transform. % without the need for the surface integral.
%%, as discussed in section \ref{Sec:LaiNorm}.
%Therefore, we may view the surface term in Eq.~(\ref{Eq:LaiNorm}) as a simple lowest order correction that can be very easily calculated.
%
%however, the term is important for the argument based on Eq.~(\ref{Eq:varIoutside}) that there is in principle a well-defined limit to the integral as $R\rightarrow\infty$.

%s and calculation domain sizes of practical interest.

\subsection{Normalization by Sauvan \emph{et al.}}
Sauvan~\emph{et al.}\cite{Sauvan_PRL_110_237401_2013} used a different approach based on the Lorentz reciprocity theorem to introduce a normalization for QNMs, in general dispersive and possibly magnetic materials, \revrev{that can be written as
\begin{align}
\langle\langle\mft_\mu|\mft_\mu\rangle\rangle_\text{Sauvan} = & \frac{1}{2}\int_V\mft_\mu(\mr)\cdot\eta(\mr,\omega)\mft_\mu(\mr)  \nonumber \\
&-\tilde{\mathbf{g}}_\mu(\mr)\cdot\kappa(\mr,\omega)\tilde{\mathbf{g}}_\mu(\mr) \ud V,
\label{Eq:SauvanNorm_org}
\end{align}
where $\tilde{\mathbf{g}}_\mu(\mr)$ is the magnetic field QNM satisfying $\nabla\times\mft_\mu(\mr)=\text{i}\tlo_\mu\mu_0\mu_\text{r}(\mr,\tlo_\mu)\tilde{\mathbf{g}}_\mu(\mr)$, $\eta(\mr,\omega) = \partial_\omega(\omega\epsilon_\text{r}(\omega,\mr))$ and $\kappa(\mr,\omega) = \partial_\omega(\omega\mu_0\mu_\text{r}(\omega,\mr))/\epsilon_0$; $\epsilon_0$ and $\mu_0$ denote the permittivity and permeability of free space, respectively, and $\mu_\text{r}(\mr,\omega)$ is the (possibly dispersive) relative permeability.} %where $\eta(\mr,\omega) = \partial_\omega(\omega\epsilon_\text{r}(\omega,\mr))$ and $\kappa(\mr,\omega) = \partial_\omega(\omega\mu_0\mu_\text{r}(\omega,\mr))/\epsilon_0$, $\epsilon_0$ and $\mu_0$ denote the permittivity and permeability of free space, respectively, $\mu_\text{r}(\mr,\omega)$ is the (possibly dispersive) relative permeability, and $\tilde{\mathbf{g}}_\mu(\mr) =\nabla\times\mft_\mu(\mr)/\text{i}\tlo_\mu\mu_0\mu_\text{r}(\mr,\tlo_\mu)$ is the magnetic field QNM.} %
This normalization seems to originate from the theory of leaky modes in waveguides\cite{Lecamp_OE_15_11048_2007}. In the original formulation and implementation in Ref.~\onlinecite{Sauvan_PRL_110_237401_2013}, there is no explicit requirement of the volume to tend to infinity. Instead, the %although this is implicitly the case, because the
integral is performed over both the usual calculation domain and the surrounding PMLs in which the real space coordinates are rotated into the complex plane, causing the propagating waves to decrease exponentially. 

For isotropic and non-magnetic materials, we may write \revrev{Eq.~(\ref{Eq:SauvanNorm_org})} %the norm 
in terms of the electric field QNMs only as %possibly the normalization
%\comm{
%\begin{align}
%\langle\langle\mft_\mu|\mft_\mu\rangle\rangle_\text{Sauvan} = & \frac{1}{2}\int_V\mft_\mu(\mr)\cdot\sigma_\epsilon(\tlo_\mu,\mr)\mft_\mu(\mr)  \nonumber \\
%&-\tilde{\mathbf{g}}_\mu(\mr)\cdot\sigma_\mu(\tlo_\mu,\mr)\tilde{\mathbf{g}}_\mu(\mr) \ud V,
%\label{Eq:SauvanNorm_org}
%\end{align}
%}
%in which
%
\begin{align}
\langle\langle\mft_\mu|\mft_\mu\rangle\rangle_\text{Sauvan} = & \frac{1}{2}\int_V\eta(\mr,\tlo_\mu)\mft_\mu(\mr)\cdot\mft_\mu(\mr) \nonumber \\
&+ \frac{1}{\tilde{k}_\mu^2} \big(\nabla\times\mft_\mu(\mr)\big)\!\cdot\!\big(\nabla\times\mft_\mu(\mr)\big) \ud V.
\label{Eq:SauvanNorm}
\end{align}
%where $\eta(\mr,\omega) = \partial_\omega(\omega\epsilon_\text{r}(\omega,\mr))$; for dispersionless materials this factor reduces to $\eta(\omega,\mr)=\epsilon_\text{r}(\mr)$. %; for non-dispersive materials  a similar generalization of the permittivity to dispersive materials in one dimension was introduced by Leung \emph{et al.} in Ref.~\onlinecite{Leung_PRA_49_3057_1994}. %where the norm is slightly rewritten,  and for non-dispersive and non-magnetic materials only, for easier comparison to Eq.~(\ref{Eq:LaiNorm}). % and (\ref{Eq:MuljarovNorm}).
\revrev{For such materials,} the correspondence with the normalization of Lai \emph{et al.} was shown in Ref.~\onlinecite{Ge_arXiv_1312.2939_2013}, and is repeated here for %the sake of
completeness. %in the case of dispersionless materials.
First, we use the vector generalization of Green's identity of the first kind~\cite{Tai_1994},
\begin{align}
\int_V (\nabla\times\mathbf{P})\cdot(\nabla\times\mathbf{Q})&- \mathbf{P}\cdot\nabla\times\nabla\times\mathbf{Q}\,\ud V   \nonumber \\
 &=\int_{\partial V}\mathbf{n}\cdot(\mathbf{P}\times\nabla\times\mathbf{Q})\,\ud A,
\end{align}
to rewrite \revrev{Eq.~(\ref{Eq:SauvanNorm})} %the integral 
as
\begin{align}
\!\langle\langle\mft_\mu|\mft_\mu\rangle\rangle_\text{Sauvan} = &\frac{1}{2}\int_V\!
\eta(\mr,\tlo_\mu)\mft_\mu(\mr)\!\cdot\mft_\mu(\mr) \nonumber\\
&\quad+\frac{1}{\tilde{k}_\mu^2}%\frac{\text{c}^2}{\tlo^2_\mu}
\mft_\mu(\mr)\!\cdot\nabla\times\nabla\times\mft_\mu(\mr)\,
\ud V \nonumber\\
&+\frac{1}{2\tilde{k}_\mu^2}%\frac{1}{2}\frac{\text{c}^2}{\tlo^2_\mu}
\int_{\partial V}\!\mathbf{n}\cdot\left(\mft_\mu(\mr)\times \nabla\times\mft_\mu(\mr) \right)\ud A.
\label{Eq:Sauvan_norm_Lai_form}
\end{align}
Then, using  the wave equation and Eq.~(\ref{Eq:SilverMullerCond}) in the limit $V\rightarrow\infty$, we recover Eq.~(\ref{Eq:LaiNorm}) with the substitution $\epsilon_\mr(\mr)\rightarrow\sigma(\mr,\tlo_\mu)$. Importantly, the integrand in Eq.~(\ref{Eq:SauvanNorm_org}) is invariant under the coordinate transformations of the PMLs~\cite{Sauvan_PRL_110_237401_2013}. Therefore, %  independent of  }In particular, therefore,
%as discussed in Ref.~\onlinecite{Ge_arXiv_1312.2939_2013},
the integral must have a well-defined value also under the trivial transformation where no coordinate rotation is performed\cite{Ge_arXiv_1312.2939_2013}. In view of Eq.~(\ref{Eq:Sauvan_norm_Lai_form}), this argument is a complement to Eq.~(\ref{Eq:varIoutside}) in suggesting that the norm has a well-defined value %there is a well-defined limit to Eq.~(\ref{Eq:LaiNorm})
as $R\rightarrow\infty$.

%The equivalence between the formulation of Lai and Sauvan, which were derived by completely different methods, suggest

%and the identity $\mathbf{A}\cdot\left(\mathbf{B}\times\mathbf{C}\right)
%= -\mathbf{B}\cdot\left(\mathbf{A}\times\mathbf{C}\right)$, we find that
%\begin{align}
%\langle\langle\mft_\text{c}|\mft_\text{c}\rangle\rangle_\text{Sauvan} &= \frac{1}{2}\int_V \epsilon_\text{r}(\mr)\mft_\text{c}\cdot\mft_\text{c}\ud V +\text{i}\frac{\sqrt{\varepsilon_\text{}}\,c}{2\tlo_\text{c}} \int_{\partial V} \mft_\text{c}\cdot\mft_\text{c}\ud A,
%%&= \langle\langle\mft_\text{c}|\mft_\text{c}\rangle\rangle_\text{Lai}
%%\int_V \epsilon_\text{r}(\mr)\mft_\text{c}\cdot\mft_\text{c}\ud\mr +\text{i}\frac{n_\text{B}c}{2\tlo_\text{c}} \int_{\partial V} \mft_\text{c}\cdot\mft_\text{c}\ud A,
%\label{Eq:SauvanEqualsLai}
%\end{align}
%and by writing explicitly , we recover Eq.~(\ref{Eq:LaiNorm}).

\subsection{Normalization by Muljarov \emph{et al.}}
Muljarov \emph{et al.}~\cite{Muljarov_arXiv_1409_6877_2014,Muljarov_EPL_92_50010_2010} derived a normalization which is similar in structure to Eq.~(\ref{Eq:LaiNorm}), but with a different surface term,
\begin{align}
\langle\langle\mft_\mu|\mft_\mu\rangle\rangle_\text{Muljarov} = &\int_V \sigma(\mr,\tlo_\mu)\mft_\mu(\mr)\cdot\mft_\mu(\mr)\ud\mr \nonumber\\
&+ \frac{\text{1}}{2\tilde{k}_\mu^2} \int_{\partial V} \mft_\mu(\mr)\cdot\partial_s\big( r \partial_r\mft_\mu(\mr)\big)\nonumber \\[1mm] &\qquad-r\big(\partial_r\mft_\mu(\mr)\big)\!\cdot\!\big(\partial_s\mft_\mu(\mr)\big)\ud A,
\label{Eq:MuljarovNorm}
\end{align}
where $\sigma(\mr,\omega)$ is defined in Eq.~(\ref{Eq:sigmaDef}), and $\partial_s$ denotes the derivative in the direction normal to the surface of the volume $V$, which needs not be spherical and needs not extend to infinity. Comparing to the two other formulas, Eq.~(\ref{Eq:MuljarovNorm}) makes explicit reference to a spherical coordinate system, the center of which one is left free to choose. % which may lead to additional interpolation induced errors when applied to
%The dependence on the coordinate system originates from the fact that the field outside the resonator may be
%appears to  %whichappears to originate with ,  originate from t
%which originates from the derivation which makes use of functional behavior
%based on the
%fact %derivation which relies on the fact
%that the field %an (implicit) expansion of the field
%outside the resonator depends on the frequency as $\tilde{k}_\mu r$, similar to the spherical hankel function. %can be written in terms of the  vector spherical harmonics.

For the case of a spherical domain of radius $R_0$, we can show the correspondence with the normalization of Lai \emph{et al.} by first evaluating the integrals $I_l^r$ in Eq.~(\ref{Eq_I_r_l}) using
%\comm{
%together with
the recurrence relations
\begin{align}
h_{l-1}(kr)=\frac{l+1}{kr}h_l(kr) + \frac{1}{k}\partial_rh_l(kr) \\
h_{l+1}(kr)=\frac{l}{kr}h_l(kr) - \frac{1}{k}\partial_rh_l(kr),
\end{align}
and the defining equation for the spherical hankel functions\revrev{, $
r^2 \partial_{rr}h_l(r) + 2r\partial_rh_l(r) + (r^2 - l(l+1))h_l(r) = 0$,} to find
%[I suggest we leave out this little calculation in the final version]}% end comm - I suggest we leave out this in the final
\begin{align}
F_l(r)&= \frac{r^2}{2\tilde{k}_\mu^2}\Big( h_l(n_\text{B}\tilde{k}_\mu r)\partial_r\big(r\partial_rh_l(n_\text{B}\tilde{k}_\mu r)\big) \nonumber \\
 &\qquad\qquad- r\big(\partial_rh_l(n_\text{B}\tilde{k}_\mu r)\big)^2 \Big).
\end{align}
Inserting in Eq.~(\ref{Eq:LaiNorm_sphericalForm}) and rearranging the terms by multiplying onto the spherical harmonics, we can write the angular integration as a surface integral in terms of the QNMs in the exact form of Eq.~(\ref{Eq:MuljarovNorm}).

%the derivations of
%Eqs.~(\ref{Eq:LaiNorm}) and (\ref{Eq:SauvanNorm}), there

%\comm{I would guess that one should be able to derive the Muljarov normalization directly from the Sauvan normalization, by manipulating the surface integral under the assumption that the field is a sum of vector spherical harmonics}

\section{Example calculations}
\label{Sec:Examples}
In this section we present a number of example calculations to illustrate that the three formulations give the same value for the norm in practical calculations. First, we consider the case of a homogeneous sphere, for which the functional form of the QNMs are known analytically and we can carry out the regularization of Eq.~(\ref{Eq:LaiNorm}) explicitly. This example was also considered by Muljarov \emph{et al.} in Ref.~\onlinecite{Muljarov_arXiv_1409_6877_2014}.
Next, we revisit the photonic crystal cavity of Ref.~\onlinecite{Kristensen_OL_37_1649_2012} and the plasmonic nanorod dimer of Ref.~\onlinecite{Ge_OL_39_4235_2014} to elaborate on the practical evaluation of Eq.~(\ref{Eq:LaiNorm}).

%The formal equivalence of Eqs.~(\ref{}) and (\ref{}) was already

%\comm{Generally on numerical calculations}

\subsection{Homogeneous sphere}
\label{Sec:Spheres}
As a first set of example applications of the three formulations, we consider QNMs of homogeneous spheres with radius $a$ in air. % homogeneous background with refractive index $n_\text{B}$. %of relative permittivity $n_\text{B}^2=n_\text{B}^2$.
Following Muljarov \emph{et al.}, we consider electric field QNMs of the transverse kind which may be written, for positions outside the sphere, in terms of the vector spherical harmonics $\mathbf{\Phi}_{lm}(\theta,\varphi)$ as
\begin{align}
\revrev{
\mft_{lm}(\mr) = h_l(\tilde{k}_\mu r)\frac{1}{\sqrt{2}}\big(\mathbf{\Phi}_{lm}(\theta,\varphi)\pm\mathbf{\Phi}^*_{lm}(\theta,\varphi)\big)
\label{Eq:transverseQNM}
}
\end{align}
with $m\geq0$, where $\mathbf{\Phi}_{lm}(\theta,\varphi)=\mr\times\nabla Y_{lm}(\theta,\varphi)$, and $Y_{lm}(\theta,\varphi)$ denote the (scalar) spherical harmonics.  %$Y_{lm}(\theta,\varphi)$ denote the spherical harmonics of order $(l,m)$ and
%\begin{align}
%R_l(kr) =\left\{ \begin{array}{cc}
%j_l(n_\text{B}kr)/j_l(ka) & \text{for } r\leq a  \\
%h_l(nkr)/h_l(ka) & \text{for } r\leq a,  \\
%\end{array} \right.
%\end{align}
%%
%in which $j_l(z)$ and $h_l(z)$ denote the spherical Bessel and Hankel functions, respectively, of order $l$.
For simplicity, we limit the discussion to the case of \revrev{purely real angular dependence (corresponding to the ``+'' in Eq.~(\ref{Eq:transverseQNM})), and} $l=1$ for which the spherical Hankel function may be written explicitly as
\begin{align}
h_1(z)=-(z+\text{i})\frac{\text{e}^{\text{i}z}}{z^2}.
\label{Eq:Hankel1}
\end{align}

Inserting in Eq.~(\ref{Eq:LaiNorm}) it follows that
\begin{align}
\langle\langle\mft_{1m}|\mft_{1m}\rangle\rangle_\text{Lai} &= I_{1m}^{r<a} + 2\left( \frac{\text{i}}{2\tilde{k}_\mu^3}-\frac{1}{\tilde{k}_\mu^4a} \right)\text{e}^{2\text{i}\tilde{k}_\mu a} \nonumber
\\
& + \lim_{R\rightarrow\infty}\left\{f_\text{Lai}^\text{res}(R)\right\},
\label{Eq:Lai_norm_spherical_l_1}
\end{align}
where
\begin{align}
I_{lm}^{r<a} &= \int_0^a\int_\Omega\epsilon_\text{r}(\mr)\mft_{lm}(\mr)\cdot\mft_{lm}(\mr)\ud V %\nonumber \\
%& \comm{= ...\text{[write out explicitly]}}
\end{align}
denotes the integral over the finite sphere, and the residual $R$ dependent factor is
\begin{align}
f_\text{Lai}^\text{res}(R) = - \frac{\text{i}}{\tilde{k}_\mu^5R^2}\text{e}^{2\text{i}\tilde{k}_\mu R}.
\label{Eq:f_res_Lai}
\end{align}
By rewriting the norm as in Eq.~(\ref{Eq_I_r_l}) we can regularize the integral and conclude that the norm is
\begin{align}
\langle\langle\mft_{1m}|\mft_{1m}\rangle\rangle_\text{Lai} &= I_{1m}^{r<a} + 2\left( \frac{\text{i}}{2\tilde{k}_\mu^3}-\frac{1}{\tilde{k}_\mu^4a} \right)\text{e}^{2\text{i}\tilde{k}_\mu a}.
\end{align}

In a similar way, we can evaluate Eq.~(\ref{Eq:Sauvan_norm_Lai_form}) to find that
\begin{align}
\langle\langle\mft_{1m}|\mft_{1m}\rangle\rangle_\text{Sauvan} &= I_{1m}^{r<a} + 2\left( \frac{\text{i}}{2\tilde{k}_\mu^3}-\frac{1}{\tilde{k}_\mu^4a} \right)\text{e}^{2\text{i}\tilde{k}_\mu a}
\nonumber \\
& + \lim_{R\rightarrow\infty}\left\{f_\text{Sauvan}^\text{res}(R)\right\},
\label{Eq:Sauvan_norm_spherical_l_1}
\end{align}
where
\begin{align}
f_\text{Sauvan}^\text{res}(R) =  \left(\frac{1}{\tilde{k}_\mu^6R^3}-\frac{\text{2i}}{\tilde{k}_\mu^5R^2}\right)\text{e}^{2\text{i}\tilde{k}_\mu R}.
\end{align}
Clearly, Eq.~(\ref{Eq:Sauvan_norm_spherical_l_1}) can be regularized in a way completely analogous to Eq.~(\ref{Eq:Lai_norm_spherical_l_1}), which is nothing but a different way of introducing the coordinate transform that was performed by use of PMLs in Ref.~\onlinecite{Sauvan_PRL_110_237401_2013}. With this approach, one finds immediately that the two norms are identical.

Finally, using %by direct application of
Eq.~(\ref{Eq:MuljarovNorm}) one can directly verify that
\begin{align}
\langle\langle\mft_{1m}|\mft_{1m}\rangle\rangle_\text{Muljarov} &= I_{1m}^{r<a} + 2\left( \frac{\text{i}}{2\tilde{k}_\mu^3}-\frac{1}{\tilde{k}_\mu^4a} \right)\text{e}^{2\text{i}\tilde{k}_\mu a},%\lim_{R\rightarrow\infty}\left\{f_\text{Lai}^\text{res}(R)\right\},
\label{Eq:Muljarov_norm_spherical_l_1}
\end{align}
so that it equals the result of the other two formulations of the norm --- notably,
without an explicit need for regularization.

\subsection{Photonic crystal cavity}
\label{Sec:pcCavity}
\begin{figure}[htb]
\flushright
\includegraphics{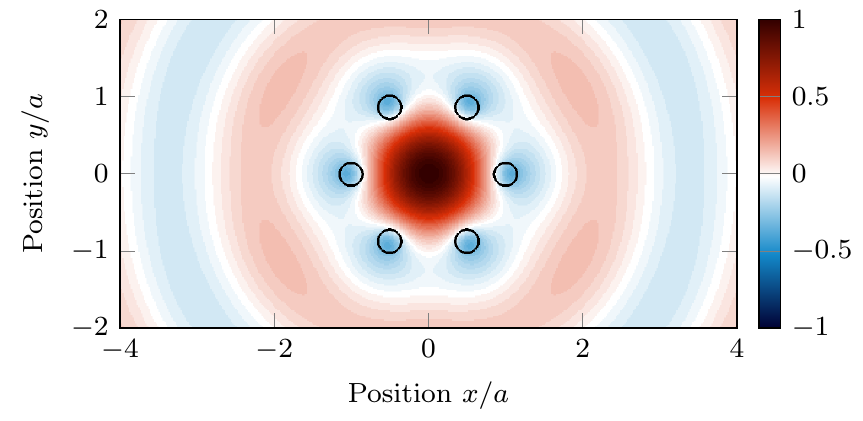}
%\beginpgfgraphicnamed{fig2}
%\begin{tikzpicture}
%\begin{axis}[
%	width=.725\columnwidth,
%	height=0.3625\columnwidth,
%	enlargelimits=false,
%	axis on top,
%	scale only axis,
%	colormap name=meep,
%	colorbar,
%	point meta min=-1,
%	point meta max=1,
%	colorbar/width=0.025\columnwidth,
%%	colorbar/height=0.6\columnwidth,
%%	colorbar style={
%%		height=0.6\columnwidth,
%%		point meta min=-1,
%%		point meta max=1,
%%		ytick={-1,-0.5,...,1}},
%	xlabel=Position $x/a$,
%	ylabel=Position $y/a$
%]
%\addplot graphics [xmin=-4,xmax=4,ymin=-2,ymax=2]
%{TM_mode_OL_real_Part_meepcolors_plus_noAxes_v3};
%\end{axis}
%\end{tikzpicture}%\\[-2.5mm]
%\endpgfgraphicnamed
\caption{\label{Fig:TM_mode_OL_real_Part_meepcolors_plus}(Color
  online) Real part of the fundamental QNM $\mft_\text{c}(\mr)$ in a small cavity made from high-index rods arranged in a hexagon with side length $a$, as indicated by black circles. The QNM is scaled to unity in the cavity center $\mr_\text{c}=(0,0)$.}
\end{figure}

As a second example, we consider the simple two dimensional photonic crystal cavity of Ref.~\onlinecite{Kristensen_OL_37_1649_2012}, which %The cavity
is formed by six cylinders of relative permittivity $\epsilon_\text{rods}=11.4$ in air. The cylinders are arranged in a hexagon with side length $a$ and have radii $r=0.15\,a$. We %. % \comm{as illustrated in XX}. We
consider the out-of-plane polarization, for which the cavity supports a QNM $\mft_\text{c}(\mr)$ with complex resonance frequency $\tlo_\text{c}a/2\pi\text{c}=0.425862- 0.013539\text{i}$, corresponding to $Q\approx16$. Figure~\ref{Fig:TM_mode_OL_real_Part_meepcolors_plus} shows the real part of $\mft_\text{c}(\mr)$ which is scaled to unity in the cavity center $\mr_\text{c}=(0,0)$.
%\begin{figure}[htb]
%%\includegraphics[width=\columnwidth]{fig1.eps}
%\flushright
%\begin{overpic}[width=.92\columnwidth]{TM_mode_OL_real_Part_meepcolors_plus_v2.eps}
%%\put(31,-5){Position $x/a$}
%%\put(13,26.5){(a)}
%%\put(-12,0){\begin{sideways}{Numerical Integral}\end{sideways}}
%\put(-7,14){\begin{sideways}{Position $y/a$}\end{sideways}}
%\put(33,-6){Position $x/a$}
%\end{overpic}\\[5mm]
%%\begin{overpic}[height=3cm]{TM_mode_OL_vQ_by_a2_complex.eps}
%%\put(39,-5){Position $x/a$}
%%\put(-4,7){\begin{sideways}{$\ft_\text{c}/\ft_\text{c}(\mr_\text{c})$}\end{sideways}}
%%\end{overpic}\qquad\;\;\\[2mm]
%\caption{\label{TM_mode_OL_real_Part_meepcolors_plus}Real part of the fundamental QNM in a small cavity made from high-index rods arranged in a hexagon with side length $a$. The QNM is scaled to unity in the cavity center $\mr_\text{c}=(0,0)$.} \end{figure}

%\input{TM_mode_OL_real_Part_meepcolors_plus.tex}

For these QNM calculations, %To calculate the cavity mode, we
we used a Fredholm type integral equation in which the QNMs appear as the self-consistent solutions to a %n integral eigenvalue equation
scattering problem with no incident field\cite{Kristensen_OL_37_1649_2012}. The equation was solved by an iterative procedure in which, for each guess of a resonance frequency $\tlo_\text{guess}$, we set up the integral equation and solved it to find the eigenvalue closest to $\tlo_\text{guess}$; the iterations then continued until the difference was smaller than some prescribed tolerance. Specifically, for each $\tlo_\text{guess}$, we discretized the integral eigenvalue equation by expanding the background Green tensor and the unknown field within each rod in a basis of cylindrical wave functions\cite{Kristensen_JOSAB_27_228_2010}. Owing to the inherent wave nature of the basis functions, each with the frequency $\tlo_\text{guess}$, this basis is particularly well suited for the expansion, and the integral formulation in terms of the Green tensor means that the radiation condition is inherently satisfied. In total, this enables very high accuracy calculations of QNMs, although the method is best suited to collections of cylindrical\cite{Kristensen_JOSAB_27_228_2010} or spherical\cite{deLasson_JOSAB_30_1996_2013} scatterers for which the projections onto the basis functions simplify significantly. %Although we used an integral equation approach for these calculations, it is worth noting that Ref.~\onlinecite{Kristensen_OL_37_1649_2012} showed explicitly that the same mode can be calculated

\begin{figure}[htb]
\flushright
\includegraphics{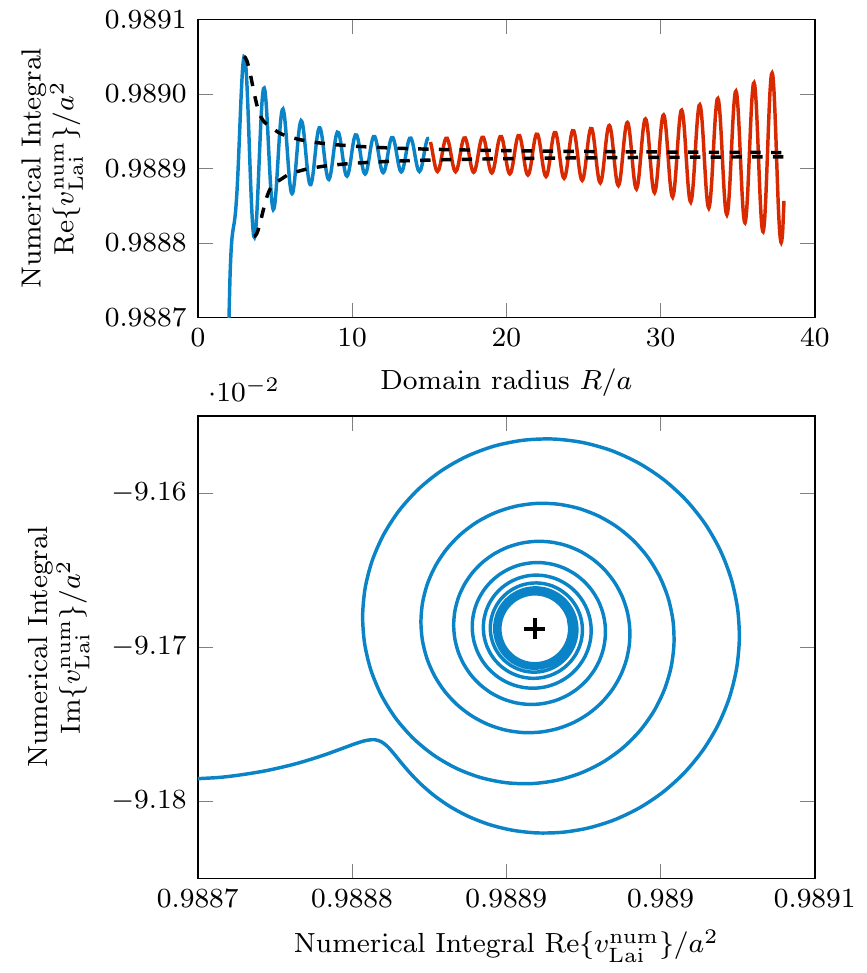}
%\beginpgfgraphicnamed{fig3}
%\begin{tikzpicture}
%  \begin{groupplot}[group style={rows=2,columns=1},
%	width=.725\columnwidth,
%    scale only axis,
%	]
%    \nextgroupplot[
%	height=.35\columnwidth,
%	smooth,
%	xmin=0,
%	xmax=40,
%	ymin=0.9887,
%	ymax=0.9891,
%	ticklabel style={
%	/pgf/number format/precision=4,
%	/pgf/number format/fixed,
%	},
%	yticklabel style={
%	/pgf/number format/fixed zerofill
%	},	
%	xlabel=Domain radius $R/a$,
%	ylabel style={align=center}, ylabel=Numerical Integral \\ $\text{Re}\{v_\text{Lai}^\text{num}\}/a^2$
%	]
%	\include{fig2a_v2}
%	\nextgroupplot[
%	smooth,
%	height=.54375\columnwidth,
%	scale only axis,
%	xmin=0.9887,
%	xmax=0.9891,
%	ymin=-0.09185,
%	ymax=-0.09155,
%	ticklabel style={
%	/pgf/number format/precision=4,
%	/pgf/number format/fixed,
%%	/pgf/number format/fixed zerofill
%	},	
%	xlabel=Numerical Integral $\text{Re}\{v_\text{Lai}^\text{num}\}/a^2$,
%	ylabel style={align=center}, ylabel=Numerical Integral \\ $\text{Im}\{v_\text{Lai}^\text{num}\}/a^2$
%	]
%	\include{fig2b}
%%	 \node[anchor=west] at (axis cs:0.98871,	-0.09158,) {(b)};
%\end{groupplot}
%\end{tikzpicture}
%\endpgfgraphicnamed
\caption{\label{Fig:TM_mode_OL_vQ_by_a2_realPart_vs_Ls}(Color
  online) Top: Real part of the calculated generalized effective mode volume as function of calculation domain radius $R$. Blue and red parts of the curve indicate the regions in which the oscillations decrease and increase in magnitude, respectively. \rev{Black dashed curves show the second order running averages starting from the first local maximum and minimum.} Bottom: Values of the calculated generalized effective mode volume as a parameterized function of calculation domain radius $R$ for the part of the curve with decreasing oscillation amplitudes, $2a<R<15a$. The center of the innermost circle corresponds to the generalized effective mode volume and is indicated with a cross.%, cf. panel (a). %. Blue and red parts of the curve indicate the regions in which the oscillations decrease and increase in magnitude, respectively, cf.
}
\end{figure}

To help in the following discussion, it is worth distinguishing between the generalized effective mode volume, defined formally in Eq.~(\ref{Eq:v_Q_def}) as a single complex number, and the numerically calculated numbers $v_i^\text{num}(R)$, with $i\in\{\text{Lai}$, $\text{Muljarov}\}$, that is obtained from Eq.~(\ref{Eq:v_Q_def}) by substituting the corresponding normalization formulas calculated in a circular domain of radius $R$. The upper panel of Fig.~\ref{Fig:TM_mode_OL_vQ_by_a2_realPart_vs_Ls} shows a zoom in at the real part of $v_\text{Lai}^\text{num}(R)$ in which an oscillatory behavior is clearly visible; from the discussion in Section~\ref{Sec:LaiNorm}, we can associate these oscillations with the residual $R$-dependent term $f_\text{Lai}^\text{res}(R)$. %, cf. Eq.~(\ref{Eq:f_res_Lai}).
A minimum in the oscillation amplitudes occurs at $R\approx15a$, after which the amplitudes increase without bounds as $R\rightarrow\infty$.

\definecolor{mycolor1}{rgb}{0.04314,0.51765,0.78039}%
\definecolor{mycolor2}{rgb}{0.84706,0.16078,0.00000}%
%
%\begin{picture}

An illustrative alternative display of the oscillatory behavior of $v_\text{Lai}^\text{num}(R)$ is found by plotting the values in the complex frequency plane as shown in the lower panel of Fig.~\ref{Fig:TM_mode_OL_vQ_by_a2_realPart_vs_Ls}.
%\begin{figure}[htb]
%%\includegraphics[width=\columnwidth]{fig1.eps}
%\flushright
%%\begin{overpic}[width=10cm]{TM_mode_OL_vQ_by_a2_realPart_vs_Ls.eps}
%%%\put(31,-5){Position $x/a$}
%%\put(-12,3){\begin{sideways}{Numerical Integral}\end{sideways}}
%%\put(-7,9){\begin{sideways}{$\text{Re}\{I_\text{Lai}\}/a^2$}\end{sideways}}
%%\put(39,-5){Domain radius $R/a$}
%%\end{overpic}\\[5mm]
%\begin{overpic}[width=0.89\columnwidth]{TM_mode_OL_vQ_by_a2_complex_intTol_1em6.eps}
%\put(40,-6){$\text{Re}\{v_\text{Lai}^\text{num}\}/a^2$}
%\put(-8,30){\begin{sideways}{$\text{Im}\{v_\text{Lai}^\text{num}\}/a^2$}\end{sideways}}
%\end{overpic}\\[5mm]
%\caption{\label{Fig:TM_mode_OL_vQ_by_a2_complex}Values of the calculated generalized effective mode volume as a parameterized function of calculation domain radius $R$. Blue and red parts of the curve indicate the regions in which the oscillations decrease and increase in magnitude, respectively, cf. Fig.~\ref{Fig:TM_mode_OL_vQ_by_a2_realPart_vs_Ls}.
%} \end{figure}
In this way one can quite easily appreciate how the values of $v_\text{Lai}^\text{num}(R)$ spirals initially towards a fixed value until the complex exponential factor in $f_\text{Lai}^\text{res}(R)$ takes over and causes the values to spiral outwards. The oscillating behavior provides a convenient and very graphical means of regularizing the integral by simply determining the center of the spiral. Clearly, this can be done with an accuracy which is much better than the smallest oscillation magnitude (better than the radius of the innermost circle in the spiral). We find that the qualitative behavior of $v_\text{Lai}^\text{num}(R)$ in the form of a spiral is not affected by the accuracy of the calculations, but the center of the spiral is.  Therefore, for all practical purposes, the error in the calculated value of $v_\text{Q}$, in this case, is dominated by the numerical integration and not the regularization of the integral.

\rev{
In practical calculations we find that one can get a convenient high-accuracy estimate of the center of the spiral by calculating the moving average
\begin{align}
v_\text{Q}^{(n)}(R_N) = \frac{1}{N}\sum_{n=n_0}^Nv_\text{Q}^{(n-1)}(R_n),
\end{align}
where $v_\text{Q}^{(0)}(R_n)$ denotes the $n$'th calculated value of $v_\text{Lai}^\text{num}(R)$. %The operation of moving averages produces curves that are smoother than the original has a smoothing effect on the curve;
In the present case of the photonic crystal cavity, the second order moving average $v_\text{Q}^{(2)}$ shows a smooth behavior for $0<R/a<40$, and by choosing two different starting points $n_0$, corresponding to consecutive local maxima and minima of the curve of $v_\text{Lai}^\text{num}(R)$, we get corresponding upper and lower bounds for $v_\text{Q}$, as illustrated in Fig.~\ref{Fig:TM_mode_OL_vQ_by_a2_realPart_vs_Ls}. % We remark, that
Because of the exponentially increasing factor in $f_\text{Lai}^\text{res}(R)$, the running averages will in general not tend to a fixed value %zero
at large distances, but may eventually start increasing (exponentially). Therefore, we do not consider their use to be a formal regularization procedure in itself, but rather a simple computational method for estimating the center of the spiral.}
%\rev{
%In practical calculations, it may be convenient to regularize the integral by appealing to the theory of divergent series\cite{Hardy_1949}. To this end, we consider each calculated value of $f_\text{Lai}^\text{res}(R_n)$ to be the $n'th$ partial sum $S_n=a_1+a_2 + ... a_n$ of a formally divergent series %$\sum_na_n$
%and evaluate $v_\text{Q}$ as the Ces\`aro sum
%\begin{align}
%v_\text{Q} = \lim_{N\rightarrow\infty}\frac{1}{N}\sum_{n=1}^Nf_\text{Lai}^\text{res}(R_n),
%\end{align}
%%\begin{align}
%%v_\text{Q} = \frac{1}{N}\sum_{R_n}f_\text{Lai}^\text{res}(R_n)
%%\end{align}
%which can be evaluated simply as the moving average of $f_\text{Lai}^\text{res}(R)$. The moving average
%}
With this procedure we find the value
%$v_\text{Q}=0.98892 - 0.09169\text{i}$.
\begin{align}
v_\text{Q}/a^2=0.988918 - 0.091688\text{i},
\label{Eq:v_Q_Lai_value}
\end{align}
with an estimated error of $|\delta v_\text{Q}|/a^2<2\times10^{-6}$.
%The qualitative behavior of $v_\text{Lai}^\text{num}(R)$ in the form of a spiral is not affected by the accuracy of the calculations, but the center of the spiral is aff
%
% spiraling behavior is independent
%
%qualitatively
%Qualitatively, the spiraling behavior appears not to be qualitatively affected by the numerical accuracy (as one would expect), but the center point does --- this too, is expected, as the center point is exactly $v_\text{Q}$.
%
%\comm{Note to Steve and Rong-Chun: Technically, I took the average of 20 points evenly distributed between $R=10a$ and $R=20a$, this can certainly be elaborated further.}
%
The oscillating behavior in Fig.~\ref{Fig:TM_mode_OL_vQ_by_a2_realPart_vs_Ls} % and \ref{Fig:TM_mode_OL_vQ_by_a2_complex}
is exactly what was noted in Ref.~\onlinecite{Kristensen_OL_37_1649_2012} which stated that ``For very low-Q cavities, however, the convergence is nontrivial due to the exponential divergence of the modes that may cause the inner product to oscillate around the proper value as a function of calculation domain size". \commMul{Importantly, the oscillations were not ``attributed to numerical issues" in this reference, as stated %by Muljarov {\em et al.}
in Ref.~\onlinecite{Muljarov_arXiv_1409_6877_2014}.} %\comm{The It is worth not
In practice, the magnitude of the oscillations are often much smaller than the desired accuracy for positions within typical calculation domains, and may also be much smaller than the overall numerical accuracy.

%In practical calculations with cavities of larger $Q$ value, the

%\comm{Compare to result without surface integral}
%\revrev{Next, we compare to results of calculations based on the formulation by Sauvan \emph{et al.} in the form of Eq.~(\ref{Eq:Sauvan_norm_Lai_form}). In practice, we first integrate over a spherical domain of radius $R_0$. Next, we change the integration contour of the remaining radial integral by setting $r=R_0+\text{i}z$ and drop the surface term, since $\mft(r,\varphi)\rightarrow0$ as $z\rightarrow0$, cf. Eq.~(\ref{Eq:HankelLimForm}). Carrying out the resulting integral, we find ...}

Next, we compare to results of calculations based on the formulation by Muljarov \emph{et al.}. Notably, in Eq.~(\ref{Eq:MuljarovNorm}) there is no requirement of an infinite calculation domain, which is very convenient from a numerical point of view. For the practical implementation, however, one must evaluate (numerically) both the first and the second derivative; together with the general accuracy of the integral, the accuracy of the derivatives then governs the overall precision. %Nevertheless, it is illustrative to evaluate also $v_\text{Muljarov}^\text{num}$ as a function of $R$ as shown together with $I_\text{Lai}$ in Fig.~\ref{Fig:TM_mode_OL_vQ_complex_Lai_and_Muljarov}. These calculations were performed using first order finite difference approximations to the derivatives with a fixed difference $\Delta r$.
Because of the finite accuracy in the calculations, % of the integral and the derivatives (as well as the integral)
there is a numerically induced phase difference between the two terms in Eq.~(\ref{Eq:MuljarovNorm}), wherefore the changes in the volume integral is not fully compensated by the surface term. This difference becomes more pronounced at larger values of $R$, because the derivatives become larger. %
Instead of investigating the behavior of Eq.~(\ref{Eq:MuljarovNorm}) as a function of calculation domain size as in Fig.~\ref{Fig:TM_mode_OL_vQ_by_a2_realPart_vs_Ls}, we %For a proper evaluation of the norm via the formulation of Muljarov \emph{et al.} we
take a fixed calculation domain size of $R=2a$ and evaluate the integrals with a fixed relative accuracy of $|\delta v_\text{Muljarov}^\text{num}|/a^2=10^{-7}$ and first order finite difference approximations to the derivatives calculated using successively smaller differences $\Delta r$. The results are shown in Table~\ref{Tab:v_Q_Muljarov} and display a convergent behavior towards a value consistent with the % the same value that was calculated using
calculation using the formulation of Lai \emph{et al.}, cf. Eq.~(\ref{Eq:v_Q_Lai_value}).
\begin{table}[htb]
\centering
\begin{tabular}{lc}
$\Delta r/a$ & $v_\text{Muljarov}^\text{num}(2a)/a^2$ \\
\hline
0.01   & 0.988894 - 0.091660\text{i}   \\ %0.988894065025352 - 0.091659875744519i
0.001  & 0.988918 - 0.091688\text{i}   \\ %0.988918216800239 - 0.091687886009183i
0.0001 & 0.988918 - 0.091688\text{i}   \\ %0.988918216800239 - 0.091687886009183i
\end{tabular}
\caption{\label{Tab:v_Q_Muljarov}Values of the calculated generalized effective mode volume $v_\text{Muljarov}^\text{num}(2a)$ for a fixed calculation domain size $R=2a$ and relative integration tolerance $\delta v_\text{Muljarov}^\text{num}/a^2=10^{-7}$ but varying accuracy of the numerical derivatives.}% in the surface term. }
\end{table}

%\comm{Note to Steve and Rong-Chun: In total, it seems that the formulation of Muljarov \emph{et al.} does in fact provide a fixed number for the integral which is equal to the result using the two other formulations --- at least to within the accuracy of these calculations.}
\revrev{Last, we compare to calculations based on the formulation by Sauvan \emph{et al.} in Eq.~(\ref{Eq:SauvanNorm_org}). Our implementation of the Fredholm integral equation does not allow for evaluation of the QNM at complex positions. Instead, we used finite-difference time-domain (FDTD) calculations\cite{Lumerical} with a run-time Fourier transform to calculate the QNM in the entire calculation domain, including the PML. The use of FDTD comes with a relatively large numerical error compared to the Fredholm integral equation approach, but has the advantage that it is immediately applicable to completely general material structures. For these calculations, we used the formulation in Eq.~(\ref{Eq:SauvanNorm_org}) based on both electric and magnetic fields, and we evaluated the components of the fields at their individual positions in the Yee cells so as to avoid interpolation errors. In principle, the evaluation of Eq.~(\ref{Eq:SauvanNorm_org}), when calculated through the PML, should be independent of the calculation domain size. In practice, however, we see a small variation with domain size, %dependence on the calculation domain size, 
which we attribute to the limited accuracy of the numerical data. %, however, we see a small dependence on the calculation domain size. 
Table~\ref{Tab:v_Q_Sauvan} summarizes the results calculated at four different domain sizes; in all cases the relative error, when compared to the value in Eq.~(\ref{Eq:v_Q_Lai_value}), is less than 2\%. %The average value is $\bar{v}^\text{num}_\text{Sauvan}=0.990 - 0.091\text{i}$ which corresponds to a relative error of less than 0.002. % when compared to the value in Eq.~(\ref{Eq:v_Q_Lai_value}). 
Finally, using the same discretization, we can also use the FDTD data to calculate the generalized effective mode volume with the formulation of Lai \emph{et al.} for which we find $v^\text{FDTD}_\text{Lai}/a^2=0.988 - 0.092\text{i}$ correspondning to a relative error of $|\delta v_\text{Q}|/|v_\text{Q}|<0.001$ when comparing to Eq.~(\ref{Eq:v_Q_Lai_value}). %with an estimated error of $|\delta v_\text{Q}|/a^2<0.001$. % four digits: $v^\text{FDTD}_\text{Lai}/a^2=0.9881 - 0.0917\text{i}$  

% note, this is really a quite conservative estimate, since from the second order running averages, the error is more on the order of 1.5e-4 for the real part and 2.4e-4 for the imaginary part. These estimates are made at Lx=10, and for larger values of Lx the oscillations show an overall shift, especially towards the negative imaginary, but this shift I think is fair to say is an artifact of the FDTD.

%, and which are all clearly in good agreement with the values calculated with Eqs.~(\ref{Eq:LaiNorm}) and (\ref{Eq:MuljarovNorm}). 

%in all cases the relative error, when compared to the value in Eq.~(\ref{Eq:v_Q_Lai_value}), is less than 2\%. To 

%In all cases the relative error, when compared to the value in Eq.~(\ref{Eq:v_Q_Lai_value}), is less than 2\%. To 

%... For the 2D FDTD calculations we used a calculation domain size of $20a\times20a$ and 

%Using 

\begin{table}[htb]
\centering
\begin{tabular}{lc}
$L/a$ & $v_\text{Sauvan}^\text{num}/a^2$ \\
\hline
6  & 0.999 - 0.091\text{i} \\ %0.9993 - 0.0911\text{i} \\ 
8  & 0.979 - 0.084\text{i} \\ %0.9789 - 0.0838\text{i} \\
10 & 0.991 - 0.102\text{i} \\ %0.9909 - 0.1021\text{i} \\
12 & 0.992 - 0.085\text{i} %0.9923 - 0.0854\text{i}
\end{tabular}
\caption{\label{Tab:v_Q_Sauvan}Values of the calculated generalized effective mode volume $v_\text{Sauvan}^\text{num}$ for different sizes of a square calculation domain with side length $L$.}
\end{table}
} 

%  \commrev{[The relative error in the Lai formulation is less than 0.001; therefore, I expect the combined FDTD calculation and numerical integration error (which I assume is comparable in the two cases) to be on this order - hence only 3 digits on the numbers in the table] }

\subsection{Plasmonic nanorod dimer}
\label{Sec:plasmonicRod}
As a final example, we %Having explored the very simple case of a 2D cavity made of circular rods, we next
%turn to a %explore a
%more complicated but typical resonator in nanophotonics and
consider the three dimensional metallic nanorod dimer in Fig.~\ref{Fig:plasmonDimer_absField}, which was recently studied in Ref.~\onlinecite{Ge_OL_39_4235_2014}. % and \onlinecite{Ge_arXiv_1312.2939_2013}.
It consists of two gold nanorods of length 100~nm and radius $r_\text{rod}=15~$nm in a homogeneous background with refractive index $n_\text{B}=1.5$. The gold nanorods are aligned along the \revrev{$y$ axis} and spaced by a gap of $20\,$nm. The relative permittivity of the gold %nanorods 
is modeled as
\begin{align}
\epsilon_\text{rod}(\omega) = 1-\frac{\omega_\text{p}^2}{\omega^2+{\rm i}\omega\gamma},
\end{align}
with $\omega_\text{p} = 1.26\times 10^{16}~$rad/s, and $\gamma = 1.41\times 10^{14}~$rad/s. The resonance frequency of the dipolar like QNM of interest is \revrev{$\tlo_\text{c}/2\pi = (291-20\text{i})$~THz, %$\tlo/2\pi = (291-\text{i}20.3)$~THz, %(291.06-\text{i}20.28)$~THz, %324.981-\rmi 16.584~$THz (1.344 - \rmi 0.0684 eV)~\cite{Ge_arXiv_1312.2939_2013},
corresponding to the wavelength $\tilde{\lambda}_\text{c}=2\pi\text{c}/\tlo = (1025+70\text{i})$~nm %$\tilde{\lambda}=2\pi\text{c}/\tlo = (1025+72\text{i})$~nm 
and a quality factor of $Q\approx 7$}.  %\comm{The corresponding mode profile is shown in Fig.~\ref{f:f4}(a). Figure~\ref{f:f4}(b) show the normalized mode along the $x$-axis with the vertical dash line indicate where the spatial divergence begins.}

%As a final example, we %Having explored the very simple case of a 2D cavity made of circular rods, we next
%%turn to a %explore a
%%more complicated but typical resonator in nanophotonics and
%consider the three dimensional metallic nanorod, which was recently studied in Refs.~\onlinecite{Sauvan_PRL_110_237401_2013} and \onlinecite{Ge_arXiv_1312.2939_2013}. Specifically, we consider a gold nanorod of length 100~nm and radius $r_c=15~$nm in a homogeneous background with $n_\text{B}^2=1.5^2$. The relative permittivity of the nanorod is $\varepsilon(\omega) = 1-\omega_p^2/(\omega^2+{\rm i}\omega\gamma)$ with $\omega_p = 1.26\times 10^{16}~$rad/s, and $\gamma = 1.41\times 10^{14}~$rad/s. For this resonator, the lowest resonance is $\tlo/2\pi = \comm{(325+16.6\text{i})}$~THz, %324.981-\rmi 16.584~$THz (1.344 - \rmi 0.0684 eV)~\cite{Ge_arXiv_1312.2939_2013},
%corresponding to the (complex) wavelength $\lambda=2\pi\text{c}/\tlo = (920 + 46.9\text{i})$~nm in agreement with Ref.~\onlinecite{Sauvan_PRL_110_237401_2013}. The Q value of the resonance is $Q\approx 10$. %\comm{The corresponding mode profile is shown in Fig.~\ref{f:f4}(a). Figure~\ref{f:f4}(b) show the normalized mode along the $x$-axis with the vertical dash line indicate where the spatial divergence begins.}

These QNM calculations were \revrev{also performed using FDTD\cite{Lumerical} with a %FDTD\cite{Lumerical}} the finite-difference time-domain (FDTD) method (Lumerical FDTD solutions\cite{Lumerical}). % with a run-time Fourier transform. %with a %finite-difference time-domain (FDTD) method with PMLs and an
%run time Fourier transform.
%and with polarization along the dimer and a , axis and a centered at  and polarized along the dimer axis, %and % symmetry axis of the nanorod is exploited, and then
%as well as a 
run-time Fourier transform based on the raw Yee cell data without interpolation,  which can be particularly detrimental in plasmonic structures due to the large field gradients. For the excitation, we used a spatial plane wave excitation near 291 THz with polarization along the dimer axis and a 6~fs Gaussian temporal profile.} %with a $46.6~$fs time window
%to obtain the spatial distribution of the QNM. %modal distribution of the QNM. of width $28~$fs to limit the data for the %select the data used to do the Fourier transform. %~\cite{Ge_OL_39_4235_2014}.
As is common practise in FDTD when analyzing metals and other high index materials, we used a non-uniform mesh consisting of \revrev{relatively small cubic Yee cells with side length $0.8$~nm along the dimer axis and 1~nm in the $x$ and $z$ directions for the inner region} %relatively small cubic Yee cells of side length 1~nm for the inner region % up to 35~nm from the surface of the nanorod,
and larger cells of side length 50 nm for the outer region. See Ref.~\onlinecite{Ge_OL_39_4235_2014} for further calculation details. Despite the fine discretization, we expect the overall accuracy of these calculations to be limited to \revrev{a few percent} %approximately 1\% 
due to meshing, sub-meshing and numerical dispersion in FDTD, which is known to be particularly demanding for metallic structures~\cite{vanVlack_2012}. We consider these to be state-of-the-art gridding parameters for metal resonators using FDTD, with simulation times on the order of days on a small computer cluster. % \comm{[maybe include specs - number of flops/sec?]}. %already take days on a supercomputer. % (so it is difficult to obtain any better spatial resolutions).

As in the case of the two-dimensional cavity, we can investigate the variation in the calculated generalized effective mode volume with calculation domain size. To this end, we define the center position $\mr_\text{c}$ as the point directly between the two nanorods and evaluate Eq.~(\ref{Eq:LaiNorm}), with the substitution $\epsilon_\text{r}(\mr)\rightarrow\sigma(\mr,\tlo_\mu)$, in a rectangular cuboid of side lengths $L_x=L_z$ and $L_y=7\,\mu$m completely enclosing the dimer. Figure~\ref{Fig:plasmonic_dimer_vQ_by_r3} shows the variation in $v_\text{Lai}^\text{num}$ as a function of domain width $L_x$. %$L_x=L_{x0}+\Delta L$ (when the height of the domain is varied accordingly as $L_y=L_{y0}+\Delta L$).
The variation in the numerical integral shows a clear oscillation as in Fig.~\ref{Fig:TM_mode_OL_real_Part_meepcolors_plus}, although the spiraling behavior is less symmetric. %, which is likely because of the lower degree of symmetry.
Despite the lower symmetry of the spiral, which we attribute to the lower degree of symmetry in the calculations, % of the cuboidal integration domain, 
\rev{we can use the second order running averages to determine the center of the oscillation with relatively high precision. With this approach, we \revrev{find %the value
%Based on the analysis in Fig. \ref{Fig:plasmonic_dimer_vQ_by_r3}, we find the value
\begin{align}
v_\text{Q}/r_\text{rod}^3 =  36.15 -0.81\text{i}. %error estimate is 0.01 
\label{Eq:V_Q_nanoRodDimer}
\end{align}
%\begin{align}
%v_\text{Q}/r^3 =  35.19 -0.96\text{i}, %35.13-3.41\text{i},
%\label{Eq:V_Q_nanoRodDimer}
%\end{align}
%
% Using a method of running averages, we find v_Q/r^3 = 35.1900 pm 0.01 - 0.9690i pm 0.0222i
%
% Using the trapz-calculated data we find v_Q/r^3 = 35.2050 pm 0.015 - 0.9704i pm 0.0078i
%
%
\definecolor{mycolor1}{rgb}{0.50196,0.50196,0.50196}%
\definecolor{mycolor2}{rgb}{0.04314,0.51765,0.78039}%
\definecolor{mycolor3}{rgb}{0.84706,0.16078,0.00000}%
\begin{figure}[t]
\flushright
\includegraphics{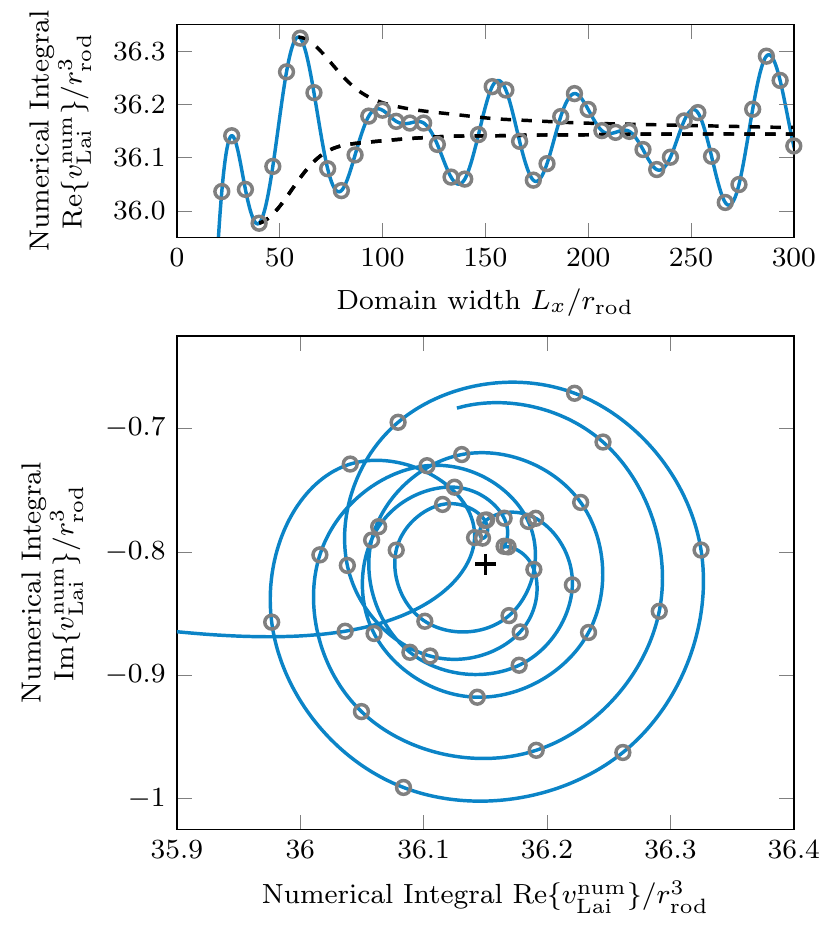}
\caption{\label{Fig:plasmonic_dimer_vQ_by_r3}(Color
  online) Top: Real part of the calculated generalized effective mode volume as function of calculation domain width $L_x$. Black dashed curves show the second order running averages. Bottom: Values of the calculated generalized effective mode volume as a parameterized function of calculation domain width. % $R$ for $L_x<300\,r_\text{rod}$.
The full curve is a spline-interpolation of the calculated data points shown by gray circles. The generalized effective mode volume $v_\text{Q}$ is indicated with a cross. %, and the estimated uncertainty is indicated with the gray shading.%, cf. panel (a). %. Blue and red parts of the curve indicate the regions in which the oscillations decrease and increase in magnitude, respectively, cf.
%\comm{[Maybe change this if new spiral-calculations by Rongchun shows a more clear spiral when varying only $L_x=L_z$ and keeping $L_y$ fixed at the maximum]}
%{\color{red}[Update figure with new data]}
}
\end{figure}
From the curves of the second order running averages, we %can give an 
estimate that the error %in Eq.~(\ref{Eq:V_Q_nanoRodDimer}) 
stemming from the regularization procedure is $|\delta v_\text{Q}|<0.01r_\text{rod}^3$. This estimate does not take into account the error in the FDTD data or the error from the numerical quadrature. As noted above, we expect that the actual uncertainty in the FDTD data may be as large as a few percent, wherefore the accuracy in $v_\text{Q}$ also in this three dimensional calculation is limited by the accuracy of the numerical calculations and not the regularization.

%As noted above, the former may be a large as a few percent, whereas we expect the latter to be ?? Therefore, the error in $v_\text{Q}$ also in this case is expected to be dominated by the error in the numerical calculations and not the regularization.

}

\revrev{
Next, we compare to calculations based on the formulation by Sauvan \emph{et al.} in Eq.~(\ref{Eq:SauvanNorm_org}). With this approach, we found $v_\text{Q}/r^3=36.17-1.02\text{i}$, corresponding to a relative difference of $|\Delta v_\text{Q}|/|v_\text{Q}|<0.006$ when comparing to Eq.~(\ref{Eq:V_Q_nanoRodDimer}) . In practise, we have found that the formulation in Eq.~(\ref{Eq:SauvanNorm_org}) is more sensitive to boundary effects in the FDTD data than Eq.~(\ref{Eq:LaiNorm}). The difference is likely caused by the material weight function being larger inside the metal resonator,  $|\sigma({\bf r},\omega)/\eta({\bf r},\omega)|\ll1$, cf. Eqs.~(\ref{Eq:LaiNorm}) and (\ref{Eq:SauvanNorm_org}). In particular, we stress that one should take care not to use interpolated field values for the integration, but rather use the raw data at the different spatial grids for the individual  components of the fields.

%, we found an (absolute) value for the generalized effective mode volume which is larger by as much as 5-10\% as compared to the (regularized) results of Eq.~(\ref{Eq:LaiNorm}). We attribute the discrepancy to the inherent Yee cell formulation of FDTD, in which the electric and magnetic fields are evaluated at different spatial points. This, in combination with the overall limited accuracy of the calculations, possibly leads to relatively larger errors when subtracting the electric and magnetic contributions to the integrand. This is a particularly sensitive point in plasmonic systems because of relatively large field gradients, and for dielectric resonators we generally find a better agreement. Also, we speculate that the data points inside the PML may lead to additional error. 

Last, we remark on our attempts of using Eq.~(\ref{Eq:MuljarovNorm}) for calculating the QNM norm of the plasmonic dimer. We have found the normalization of Muljarov \emph{et al.} very difficult to evaluate with the FDTD data to a precision in which it provides better accuracy than Eq.~(\ref{Eq:LaiNorm}). We believe that these problems are caused by the finite accuracy of the QNM calculation, which in practice is an unfortunate but rather general characteristic of many numerical Maxwell solvers in three dimensions. For FDTD, the inherent formulation in terms of Cartesian coordinates leads to additional interpolation errors when performing the radial differentiation as dictated by Eq.~(\ref{Eq:MuljarovNorm}). In the present case of the plasmonic dimer, %in particular, 
we found that the field gradients lead to relatively large numerical errors in the surface integral which in practice dominates the achievable accuracy. From the analysis in Section \ref{Sec:normIntegrals}, it is obvious that these limitations are not principal in nature, but rather practical issues pertaining to the numerical calculation method.
}

\takeout{

This estimate, which is indicated by the gray shaded areas in Fig.~\ref{Fig:plasmonic_dimer_vQ_by_r3}, is more conservative than what would follow directly from the % is calculated from the distance between the
curves of the second order running averages  because of the numerical quadrature and the lesser degree of symmetry. Nevertheless, the error is still much smaller} %, we choose a more conservative estimate and set $|\delta v_\text{Q}|<0.1r^3$}
%This error is much smaller %it is arguably possible to determine the center as the point between the smallest oscillation amplitudes for both real and imaginary part in the initial part of the curve % with an accuracy %which is better than $\delta v_\text{Q}=0.1r^3$. In pr
%which is at l as the point between the
% by the average of the
%with an accuracy
%which is at least as good as the oscillation amplitudes in the initial part of the curve
%(indicated by blue).
%With a conservative estimate we set the uncertainty to $0.1r^3$; this is better

\takeout{
than the expected overall numerical accuracy in the FDTD calculations, wherefore the error in the calculated $v_\text{Q}$ also in this case is expected to be dominated by the error in the numerical calculations and not the regularization.

}
}

%
%

%With this approach, we find the value
%%Based on the analysis in Fig. \ref{Fig:plasmonic_dimer_vQ_by_r3}, we find the value
%\begin{align}
%v_\text{Q}/r^3 =  35.19 -0.96\text{i}, %35.13-3.41\text{i},
%\label{Eq:V_Q_nanoRodDimer}
%\end{align}
%%
%% Using a method of running averages, we find v_Q/r^3 = 35.1900 pm 0.01 - 0.9690i pm 0.0222i
%%
%% Using the trapz-calculated data we find v_Q/r^3 = 35.2050 pm 0.015 - 0.9704i pm 0.0078i
%%
%%
%with an estimated error of $|\delta v_\text{Q}|<0.1r^3$,
%circle in the lower panel of Fig.~\ref{Fig:plasmonic_dimer_vQ_by_r3}.

\takeout{

The generalized effective mode volume in Eq.~(\ref{Eq:V_Q_nanoRodDimer}) leads to \revrev{a relative} enhanced spontaneous emission rate %$\Gamma$, relative to the rate in the homogeneous background, $\Gamma_\text{B}$, 
of %a Purcell factor of
$\Gamma/\Gamma_\text{B}=1475\pm8$, as calculated using \revrev{Eq.~(\ref{Eq:F_P_from_G}) and} a single QNM approximation to the Green tensor\cite{Kristensen_ACSphot_1_2_2014}. %; \comm{[I did not include the factor of +1, since the position $\mr_\text{c}$ in this case is really inside the structure]} \rev{because of the low $Q$ value, direct application of the Purcell formula in this case gives the markedly different result of $F_\text{P}'=1485$}.
The stated uncertainty is on the order of 0.5\% and is a simple estimate calculated as the difference in the extremal Purcell factors resulting from the finite accuracy of $v_\text{Q}$. %by inserting $v_\text{Q}$ from Eq. a crude estimate from using  in the Purcell formula.}
Assuming negligible quasi-static coupling, we can assess this value by comparing to independent \revTwo{reference calculations of the spontaneous emission rate} %Purcell factor
using FDTD to determine the electric field Green tensor. With this approach we find \revTwo{$\Gamma^\text{ref}/\Gamma_\text{B}=1484\pm2$}. These reference calculations were performed with the same mesh as the QNM calculation, wherefore we expect the overall intrinsic numerical error due to discretization to be less important for the comparison. %, although for comparing to calculations using other methods we would comparison.
The two independent calculations agree to within the estimated errors; notably, the QNM approach predicts a value that is lower than the reference calculation, which is consistent with the possibility of non-vanishing quasi-static coupling due to the proximity of $\mr_\text{c}$ to the metal surfaces. %\comm{[Note, this also includes the factor of 1]} % Notably, for this comparison we expect the intrinsic numerical error of the calculations due to the discretization %, however, derive from the discretization of the equation system and so to be %are expected to be
%the same for both the QNM calculation and the reference calculations which were performed on the same mesh.

%\comm{[Maybe insert here discussions of results using the Sauvan formulation. This would really be a good addition to the story, since then we compare Lai to Muljarov in III B and Lai to Sauvan in III C]}

%\rev{For the calculations on the plasmonic nanorod dimer, we have been unable to }
%We believe that this may be caused by the

%\revTwo{We believe that these problems are likely caused by the finite accuracy of numerical maxwell solvers in three dimensions, such as FDTD}

%For dielectric resonators (not shown) we have found better agreement with the (regularized) results of Eq.~(\ref{})

} %takeout

\subsection*{Discussion}
Comparing Eqs.~(\ref{Eq:LaiNorm}), (\ref{Eq:Sauvan_norm_Lai_form}) and (\ref{Eq:MuljarovNorm}), % \rev{it is tempting to view} %
we can understand
the three formulations of the norm as arising from three different approaches to the regularization of the integral
\begin{align}
\int\epsilon_\text{r}(\mr)\mft_\mu(\mr)\cdot\mft_\mu(\mr)\ud V.
\end{align}
This integral can be regularized immediately by a complex coordinate transform, as discussed in section \ref{Sec:LaiNorm}. Therefore, in view of Eqs.~(\ref{Eq:MuljarovNorm}) one may view the surface term in Eq.~(\ref{Eq:LaiNorm}) as a \rev{simple choice} of %, \comm{lowest order}
regularization for use with \rev{data at positions along the real axis}. It is worth noting, however, that the surface term appears directly in perturbation theory calculations\cite{Lai_PRA_41_5187_1990}, or from Eq.~(\ref{Eq:Sauvan_norm_Lai_form}), under the assumption of the Silver-M\"uller radiation condition, and it is in principle important for the argument based on Eqs.~(\ref{Eq:varIoutside}) or~(\ref{Eq:Sauvan_norm_Lai_form}) that one can assign a well-defined value to the norm as $R\rightarrow\infty$. In many  practical calculations %, such as for the plasmonic dimer in Fig.~\ref{Fig:plasmonDimer_absField}, %as we illustrate below,
we find that the relatively simple formula of Lai \emph{et al.} is very often sufficient, and that the uncertainty in the norm (or the associated effective mode volume) is dominated by the %typically derives from
uncertainty in the numerical data. When one has access to complex position data, the normalization of Sauvan \emph{et al.} represents an obvious alternative, and for high-accuracy numerical data, or when the QNMs are known analytically, one can benefit the most from the formula of Muljarov \emph{et al.}.

In this Article, we have focused on normalization of resonators embedded in a homogeneous background material, for which \rev{we have argued that} Eq.~(\ref{Eq:SilverMullerCond}) represents the proper choice of radiation condition. For resonators coupled to waveguides the radiation condition is different, but also in this case there appears to be an inherent need for regularization of the normalization integral\cite{Kristensen_OL_39_6359_1014}. Also, it is worth noting that Bai \emph{et al.} recently introduced an alternative calculation method for QNMs, based on a scattering formulation at complex frequencies, in which the resulting QNMs are naturally normalized without an explicit need for an integration step\cite{Bai_OE_21_27371_2013}.

%\rev{[Maybe also point out that all these calculations were performed for super-low Q caivities, and that in general there is really no problem]}

% fa mthe FDTD calculation: from the center of the nanorod to 10~nm away from the surface along the $x$-axis we used a mesh size of 1~nm, after which a mesh size of 40~nm has been employed. These are state-of-the-art gridding parameters for metal resonators using FDTD. % and simulation times already take days on a supercomputer. % (so it is difficult to obtain any better spatial resolutions).

\section{Conclusions}
\label{Sec:Conclusions}
In conclusion, we have discussed the relation between three different formulations for the normalization of QNMs in leaky optical cavities and plasmonic resonators. The three formulations arise from independent derivations of an expansion for the electromagnetic field in terms of QNMs, and for %we have shown how one can
spherical calculation volumes, we have shown the equivalence between the different formulas explicitly. \revrev{Moreover, we have illustrated the equivalence between the norms with a number of example calculations of practical interest.} \rev{We have argued, that the} apparent differences in the normalization formulas \rev{can be understood as arising} %arise
from alternative ways of regularizing the normalization integral to handle the inherently divergent QNMs. In this view, the surface term in the normalization by Lai \emph{et al.} \rev{represents a simple choice of regularization} %\comm{lowest order} correction
which in principle is insufficient to properly regularize the integral when the size of the calculation domain is varied along the real axis. We have discussed this issue in detail, and we have shown how one can, in principle, always regularize the integral by a complex coordinate transform. In practice, we find that there is rarely any need for additional regularization beyond the relatively simple formula of Lai \emph{et al.} \rev{Nevertheless, depending on the calculation method used to obtain the QNMs, the normalizations of Sauvan \emph{et al.} or Muljarov \emph{et al.} may be \rev{more convenient.} %  faster or more accurate.  %We regard the three formulations as being complementary, and
Regardless of the choice of normalization integral, we have discussed how one can use the norm to define an effective mode volume for use in the Purcell formula. Because of the complementarity of the three norms, this naturally leads to the same estimates of the enhanced spontaneous emission factor.}
\\

%We discuss three formally different formulas for %elaborate on the various
%normalization of %formulas for
%quasinormal modes currently in %current
%use for modeling optical cavities and plasmonic resonators and show that they are complementary and provide the same result.

This work was supported by the Carlsberg foundation, Queen's University and the Natural Sciences and Engineering Research Council of Canada.

%\bibliography{normalization_bib}

\begin{thebibliography}{10}
\expandafter\ifx\csname url\endcsname\relax
  \def\url#1{\texttt{#1}}\fi
\expandafter\ifx\csname urlprefix\endcsname\relax\def\urlprefix{URL }\fi
\providecommand{\bibinfo}[2]{#2}
\providecommand{\eprint}[2][]{\url{#2}}

\bibitem{Purcell_PR_69_681_1946}
\bibinfo{author}{Purcell, E.~M.}
\newblock \emph{Spontaneous Emission Probabilities at Radio Frequencies}.
\newblock \bibinfo{journal}{Physical Review} \textbf{\bibinfo{volume}{69}},
  \bibinfo{pages}{681} (\bibinfo{year}{1946}).

\bibitem{Ching_1996}
\bibinfo{author}{Ching, E. S.~C.}, \bibinfo{author}{Leung, P.~T.} \&
  \bibinfo{author}{Young, K.}
\newblock \emph{\bibinfo{title}{Optical processes in microcavities---the role
  of quasi-normal modes, R. K. Chang and A. J. Campillo Eds.}}
  (\bibinfo{publisher}{World Scientic}, \bibinfo{year}{1996}).

\bibitem{Ching_RevModPhys_70_1545_1998}
\bibinfo{author}{Ching, E. S.~C.}, \bibinfo{author}{Leung, P.~T.},
  \bibinfo{author}{Maassen van~den Brink, A.}, \bibinfo{author}{Suen, W.~M.},
  \bibinfo{author}{Tong, S.~S.} \& \bibinfo{author}{Young, K.}
\newblock \emph{Quasinormal-mode expansion for waves in open systems}.
\newblock \bibinfo{journal}{Review of Modern Physics}
  \textbf{\bibinfo{volume}{70}}, \bibinfo{pages}{1545--1554}
  (\bibinfo{year}{1998}).

\bibitem{Kristensen_ACSphot_1_2_2014}
\bibinfo{author}{Kristensen, P.~T.} \& \bibinfo{author}{Hughes, S.}
\newblock \emph{Modes and Mode Volumes of Leaky Optical Cavities and Plasmonic
  Nanoresonators}.
\newblock \bibinfo{journal}{ACS Photonics} \textbf{\bibinfo{volume}{1}},
  \bibinfo{pages}{2--10} (\bibinfo{year}{2014}).

\bibitem{Kristensen_OL_37_1649_2012}
\bibinfo{author}{Kristensen, P.~T.}, \bibinfo{author}{Vlack, C.~V.} \&
  \bibinfo{author}{Hughes, S.}
\newblock \emph{Generalized effective mode volume for leaky optical cavities}.
\newblock \bibinfo{journal}{Optics Letters} \textbf{\bibinfo{volume}{37}},
  \bibinfo{pages}{1649--1651} (\bibinfo{year}{2012}).

\bibitem{Lai_PRA_41_5187_1990}
\bibinfo{author}{Lai, H.~M.}, \bibinfo{author}{Leung, P.~T.},
  \bibinfo{author}{Young, K.}, \bibinfo{author}{Barber, P.~W.} \&
  \bibinfo{author}{Hill, S.~C.}
\newblock \emph{Time-independent perturbation for leaking electromagnetic modes
  in open systems with application to resonances in microdroplets}.
\newblock \bibinfo{journal}{Physical Review A} \textbf{\bibinfo{volume}{41}},
  \bibinfo{pages}{5187--5198} (\bibinfo{year}{1990}).

\bibitem{Leung_PRA_49_3057_1994}
\bibinfo{author}{Leung, P.~T.}, \bibinfo{author}{Liu, S.~Y.} \&
  \bibinfo{author}{Young, K.}
\newblock \emph{Completeness and time-independent perturbation of the quasinormal modes
of an absorptive and leaky cavity}.
\newblock \bibinfo{journal}{Physical Review A} \textbf{\bibinfo{volume}{49}},
  \bibinfo{pages}{3982--3989} (\bibinfo{year}{1994}).

\bibitem{Leung_JOSAB_13_805_1996}
\bibinfo{author}{Leung, P.~T.} \& \bibinfo{author}{Pang, K.~M.}
\newblock \emph{Completeness and time-independent perturbation of
  morphology-dependent resonances in dielectric spheres}.
\newblock \bibinfo{journal}{Journal of the Optical Society of America B}
  \textbf{\bibinfo{volume}{13}}, \bibinfo{pages}{805} (\bibinfo{year}{1996}).

\bibitem{Lee_JOSAB_16_1409_1999}
\bibinfo{author}{Lee, K.~M.}, \bibinfo{author}{Leung, P.~T.} \&
  \bibinfo{author}{Pang, K.~M.}
\newblock \emph{Dyadic formulation of morphology-dependent resonances. I.
  Completeness relation}.
\newblock \bibinfo{journal}{Journal of the Optical Society of America B}
  \textbf{\bibinfo{volume}{16}}, \bibinfo{pages}{1409--1417}
  (\bibinfo{year}{1999}).

\bibitem{Lee_JOSAB_16_1418_1999}
\bibinfo{author}{Lee, K.~M.}, \bibinfo{author}{Leung, P.~T.} \&
  \bibinfo{author}{Pang, K.~M.}
\newblock \emph{Dyadic formulation of morphology-dependent resonances. II.
  Perturbation theory}.
\newblock \bibinfo{journal}{Journal of the Optical Society of America B}
  \textbf{\bibinfo{volume}{16}}, \bibinfo{pages}{1418--1430}
  (\bibinfo{year}{1999}).

\bibitem{Sauvan_PRL_110_237401_2013}
\bibinfo{author}{Sauvan, C.}, \bibinfo{author}{Hugonin, J.~P.},
  \bibinfo{author}{Maksymov, I.~S.} \& \bibinfo{author}{Lalanne, P.}
\newblock \emph{Theory of the Spontaneous Optical Emission of Nanosize Photonic
  and Plasmon Resonators}.
\newblock \bibinfo{journal}{Physical Review Letters}
  \textbf{\bibinfo{volume}{110}}, \bibinfo{pages}{237401}
  (\bibinfo{year}{2013}).

\bibitem{Muljarov_arXiv_1409_6877_2014}
\bibinfo{author}{Muljarov, E.~A.}, \bibinfo{author}{Doost, M.~B.} \&
  \bibinfo{author}{Langbein, W.}
\newblock \emph{Exact mode volume and Purcell factor of open optical systems}.
\newblock \bibinfo{journal}{arXiv:1409.6877 [cond-mat-mes.hall]}
  (\bibinfo{year}{2014}).

\bibitem{Muljarov_EPL_92_50010_2010}
\bibinfo{author}{Muljarov, E.~A.}, \bibinfo{author}{Langbein, W.} \&
  \bibinfo{author}{Zimmermann, R.}
\newblock \emph{Brillouin-Wigner perturbation theory in open electromagnetic
  systems}.
\newblock \bibinfo{journal}{European Physics Letters}
  \textbf{\bibinfo{volume}{92}}, \bibinfo{pages}{50010} (\bibinfo{year}{2010}).

\bibitem{Martin_MultipleScattering}
\bibinfo{author}{Martin, P.}
\newblock \emph{\bibinfo{title}{Multiple Scattering. Interaction of
  time-harmonic waves with N obstacles}} (\bibinfo{publisher}{Cambridge
  University Press}, \bibinfo{year}{2006}).

\bibitem{Koenderink_OL_35_4208_2010}
\bibinfo{author}{Koenderink, A.~F.}
\newblock \emph{On the use of Purcell factors for plasmon antennas}.
\newblock \bibinfo{journal}{Optics Letters} \textbf{\bibinfo{volume}{35}},
  \bibinfo{pages}{4208--4210} (\bibinfo{year}{2010}).

\bibitem{Ge_arXiv_1312.2939_2013}
\bibinfo{author}{Ge, R.-C.}, \bibinfo{author}{Kristensen, P.~T.},
  \bibinfo{author}{Young, J.~F.} \& \bibinfo{author}{Hughes, S.}
\newblock \emph{Quasinormal mode approach to modelling light-emission and
  propagation in nanoplasmonics}.
\newblock \bibinfo{journal}{New Journal of Physics}
  \textbf{\bibinfo{volume}{16}}, \bibinfo{pages}{113048}
  (\bibinfo{year}{2014}).

\bibitem{Zeldovich_SPJ_12_542_1961}
\bibinfo{author}{Zel'dovich, Y.~B.}
\newblock \emph{On the theory of unstable states}.
\newblock \bibinfo{journal}{Soviet Physics JETP} \textbf{\bibinfo{volume}{12}},
  \bibinfo{pages}{542} (\bibinfo{year}{1961}).

\bibitem{Barrera_EJP_6_287_1985}
\bibinfo{author}{Barrera, R.~G.}, \bibinfo{author}{Est\'{e}vez, G.~A.} \&
  \bibinfo{author}{Giraldo, J.}
\newblock \emph{Vector spherical harmonics and their application to
  magnetostatics}.
\newblock \bibinfo{journal}{European Journal of Physics}
  \textbf{\bibinfo{volume}{6}}, \bibinfo{pages}{287} (\bibinfo{year}{1985}).

\bibitem{SnyderLoveBook}
\bibinfo{author}{Snyder, A.~W.} \& \bibinfo{author}{Love, J.}
\newblock \emph{\bibinfo{title}{Optical Waveguide Theory}}
  (\bibinfo{publisher}{Chapman and Hall}, \bibinfo{year}{1983}).

\bibitem{Lecamp_OE_15_11048_2007}
\bibinfo{author}{Lecamp, G.}, \bibinfo{author}{Hugonin, J.~P.} \&
  \bibinfo{author}{Lalanne, P.}
\newblock \emph{Theoretical and computational concepts for periodic optical
  waveguides}.
\newblock \bibinfo{journal}{Optics Express} \textbf{\bibinfo{volume}{15}},
  \bibinfo{pages}{11042--11060} (\bibinfo{year}{2007}).

\bibitem{Tai_1994}
\bibinfo{author}{Tai, C.-T.}
\newblock \emph{\bibinfo{title}{Dyadic Green Functions in Electromagnetic
  Theory, 2nd ed.}} (\bibinfo{publisher}{IEEE Press}, \bibinfo{year}{1994}).

\bibitem{Ge_OL_39_4235_2014}
\bibinfo{author}{Ge, R.-C.} \& \bibinfo{author}{Hughes, S.}
\newblock \emph{Design of an efficient single photon source from a metallic
  nanorod dimer: a quasi-normal mode finite-difference time-domain approach}.
\newblock \bibinfo{journal}{Optics Letters} \textbf{\bibinfo{volume}{39}},
  \bibinfo{pages}{4235} (\bibinfo{year}{2014}).

\bibitem{Kristensen_JOSAB_27_228_2010}
\bibinfo{author}{Kristensen, P.~T.}, \bibinfo{author}{Lodahl, P.} \&
  \bibinfo{author}{M{\o}rk, J.}
\newblock \emph{Light propagation in finite-sized photonic crystals: Multiple
  scattering using an electric field integral equation}.
\newblock \bibinfo{journal}{Journal of the Optical Society of America B}
  \textbf{\bibinfo{volume}{27}}, \bibinfo{pages}{228--237}
  (\bibinfo{year}{2010}).

\bibitem{deLasson_JOSAB_30_1996_2013}
\bibinfo{author}{de~Lasson, J.~R.}, \bibinfo{author}{M{\o}rk, J.} \&
  \bibinfo{author}{Kristensen, P.~T.}
\newblock \emph{Three-dimensional integral equation approach to light
  scattering, extinction cross sections, local density of states, and
  quasi-normal modes}.
\newblock \bibinfo{journal}{Journal of the Optical Society of America B}
  \textbf{\bibinfo{volume}{30}}, \bibinfo{pages}{1996--2007}
  (\bibinfo{year}{2013}).

\bibitem{Lumerical}
\bibinfo{note}{We used FDTD solutions V. 8.6.3, www.lumerical.com}.

\bibitem{vanVlack_2012}
\bibinfo{author}{Vlack, C.~V.}
\newblock \emph{\bibinfo{title}{Dyadic Green Functions and their applications
  in Classical and Quantum Nanophotonics}}.
\newblock Ph.D. thesis, \bibinfo{school}{Queen's University}
  (\bibinfo{year}{2012}).

\bibitem{Kristensen_OL_39_6359_1014}
\bibinfo{author}{Kristensen, P.~T.}, \bibinfo{author}{de~Lasson, J.~R.} \&
  \bibinfo{author}{Gregersen, N.}
\newblock \emph{Calculation, normalization, and perturbation of quasinormal
  modes in coupled cavity-waveguide systems}.
\newblock \bibinfo{journal}{Optics Letters} \textbf{\bibinfo{volume}{39}},
  \bibinfo{pages}{6359} (\bibinfo{year}{2014}).

\bibitem{Bai_OE_21_27371_2013}
\bibinfo{author}{Bai, Q.}, \bibinfo{author}{Perrin, M.},
  \bibinfo{author}{Sauvan, C.}, \bibinfo{author}{Hugonin, J.-P.} \&
  \bibinfo{author}{Lalanne, P.}
\newblock \emph{Efficient and intuitive method for the analysis of light
  scattering by a resonant nanostructure}.
\newblock \bibinfo{journal}{Optics Express} \textbf{\bibinfo{volume}{21}},
  \bibinfo{pages}{27371--27382} (\bibinfo{year}{2013}).

\end{thebibliography}

\end{document}